\documentclass[aps,twocolumn,floatfix]{revtex4-1}
\usepackage{times}
\usepackage{graphics}
\usepackage{grffile} 
\usepackage{epsfig,calc,xcolor,xspace}

\usepackage{bm}

\usepackage{placeins} 
\usepackage[breaklinks=true,
            pdfborder={0 0 1},
            colorlinks=true,
            linkcolor=black,
            citecolor=blue,
            urlcolor=blue]{hyperref}

\usepackage{amssymb,amsmath,amsbsy,wasysym}
\usepackage{chemarr} 
\usepackage[normalem]{ulem}
\usepackage{verbatim}  

\definecolor{dpurple}{rgb}{0.5,0.0,0.9}
\definecolor{dlgreen}{rgb}{0.0,0.7,0.0}

\newcommand{\ie}{\textit{i.e.}\xspace}

\newcommand{\etal}{et al.\xspace}
\newcommand{\kt}{k_{\text{B}}T}
\newcommand{\egg}{\epsilon_{\text{gg}}}
\newcommand{\eng}{\epsilon_{\text{ng}}}
\newcommand{\kmem}{\kappa_{\text{mem}}}
\newcommand{\kcap}{\kappa_{\text{shell}}}
\newcommand{\dhead}{d_{\text{head}}}
\newcommand{\Gshell}{{G_n^{\text{shell}}}}

\newcommand{\re}{r_{\text{e}}}
\newcommand{\reqi}{r^{\text{eq}}_{i}}
\newcommand{\al}{\alpha_{l}}
\newcommand{\li}{l_{i}}
\newcommand{\UMi}{U^{\text{M}}_{i}}
\newcommand{\UggM}{U_{\text{gg}}^{\text{M}}}

\def\btt1{{\tt$\backslash$\string1}}%

\def\AmS{{\protect\the\textfont2
        A\kern-.1667em\lower.5ex\hbox{M}\kern-.125emS}}

\citeindextrue

\begin{document}

\title{Why enveloped viruses need cores -- the contribution of a nucleocapsid core to viral budding}
\author{Guillermo R. L\'azaro$^1$}
\author{Suchetana Mukhopadhyay$^2$}
\author{Michael F. Hagan$^1$}\email{hagan@brandeis.edu}
\affiliation{$^1$ Martin Fisher School of Physics, Brandeis University, Waltham, MA, 02454, USA\\
$^2$ Department of Biology, Indiana University, 212 S. Hawthorne Drive, Bloomington, IN 47405, USA}
\date{\today}
\begin{abstract}
%
During the alphavirus lifecycle, a nucleocapsid core buds through the cell membrane to acquire an outer envelope of lipid membrane and viral glycoproteins. However, the presence of a nucleocapsid core is not required for assembly of infectious particles.  To determine the role of the nucleocapsid core, we develop a coarse-grained computational model with which we investigate budding dynamics as a function of glycoprotein and nucleocapsid interactions, as well as budding in the absence of a nucleocapsid. We find that there is a transition between glycoprotein-directed budding and nucleocapsid-directed budding which occurs above a threshold strength of nucleocapsid interactions.
The simulations predict that  glycoprotein-directed budding leads to significantly increased size polydispersity and particle polymorphism. This polydispersity can be qualitatively explained by a theoretical model accounting for the competition between bending energy of the membrane and the glycoprotein shell. The simulations also show that the geometry of a budding particle leads to a barrier to subunit diffusion, which can result in a stalled, partially budded state. We present a phase diagram for this and other morphologies of budded particles.  Comparison of these structures against experiments
could establish bounds on whether budding is directed by glycoprotein or nucleocapsid interactions. Although our model is motivated by alphaviruses, we discuss implications of our results for other enveloped viruses.
\end{abstract}

\maketitle


\section{Introduction}
\label{sec:intro}
Membrane remodeling is required for critical cellular processes including endocytosis, formation of multivesicular bodies, retrograde trafficking and exosome formation. Viruses and other pathogens also reshape cellular membranes during different stages of their lifecycles including entry into the host cell, formation of replication complexes, construction of assembly factories, and exit (also called egress or budding). Understanding the mechanisms of viral budding and the forces that drive this process would advance our fundamental understanding of viral lifecycles, and shed light on analogous cellular processes in which membrane remodeling and vesicle formation are essential for function.
In parallel, understanding fundamental determinants of budding and membrane dynamics would facilitate the design of viral nanoparticles.  There is keen interest in reengineering enveloped viral nanoparticles to be used as targeted transport vehicles capable of crossing cell membranes through viral fusion \cite{Lundstrom2009,Cheng2013,Rowan2010,Petry2003,Rohovie2017}.

All viruses contain a capsid protein, which primarily functions to protect the viral genome during viral transmission. In enveloped viruses, the internal capsid is surrounded by a host-derived lipid bilayer and viral glycoproteins (GPs) embedded in this membrane.  Enveloped viruses can be sub-divided into two groups based on their sequence of virion assembly and budding.  For the first group (e.g. alphaviruses, hepatitis B, herpes) budding requires the assembly of a preformed nucleocapsid core (NC), which may be ordered or disordered depending on the virus. The core then binds to membrane-bound GPs and initiates budding \cite{Sundquist2012,Hurley2010,Welsch2007,Garoff1998}. For the second group, (e.g. influenza, type C retroviruses (HIV)) capsid assembly occurs concomitant with budding \cite{Welsch2007,Sundquist2012,Vennema1996}. The advantage of one assembly mechanism over another is not obvious; particle infectivity, morphology, and stability may all influence the preferred budding process.

The importance of preformed capsids in Alphavirus assembly is of particular interest because the presence of a capsid in the particle is not necessary for production of infectious particles. The traditional view is that alphaviruses follow the preassembled NC budding pathway \cite{Garoff1978,Garoff2004,Strauss1994,Wilkinson2005}, based on the observation of high concentrations of NCs in the cytoplasm \cite{Acheson1967}, and evidence that GP-GP and NC-GP interactions are required for virion formation \cite{Suomalainen1992,Lopez1994}. However, several studies have challenged this conclusion. In particular, Forsell \etal \cite{Forsell2000} reported successful assembly and budding of alphavirus despite mutations which inhibited NC assembly by impairing interactions between NC proteins, while Ruiz-Guillen \etal \cite{Ruiz-Guillen2016} observed budding of infectious alphavirus particles from cells which did not express the capsid gene. In both cases infectious particles were assembled and released or budded from the cell.  These observations suggest that GP interactions may be sufficient for alphavirus budding. This begs the question: Why do enveloped viruses have internal nucleocapsid cores? Is there an advantage to having a NC during budding?

Molecular dynamics (MD) simulations can be a useful tool to bridge the gaps between the different steps of assembly that cannot be experimentally characterized. Computational studies have already provided insightful information about virus NC assembly \cite{Perlmutter2015}, as well as the interactions between proteins and lipid membranes \cite{Reynwar2007,Simunovic2013,Bradley2016}. Previous simulations on budding of nanoscale particles led to important insights but did not consider the effect of GPs (\cite{Smith2007,Vacha2011,Ruiz-Herrero2012,Deserno2002,Jiang2015,Li2010a, Li2010,Yang2011}), although budding directed by GP adsorption or capsid assembly have been the subject of continuum theoretical modeling \cite{Lerner1993,Tzlil2004,Zhang2008}. The formation of clathrin cages during vesicle secretion, a process which bears similarities to viral budding, has also been the subject of modeling studies \cite{Foret2008,Cordella2014, Cordella2015,Matthews2013}. Most closely related to our work are previous simulations on the assembly and budding of 12-subunit capsids, which found that membrane adsorption can lower entropic barriers to assembly \cite{Matthews2012,Matthews2013a} and that membrane microdomains can facilitate assembly and budding \cite{Ruiz-Herrero2015}. In contrast to these earlier works, we consider the presence of a nucleocapsid, a larger shell (80 trimer subunits), and a different subunit geometry. We find that these modifications lead to qualitatively different assembly pathways and outcomes in some parameter ranges.

\begin{figure}[hbt]
  \begin{center}
  \includegraphics[width=\columnwidth]{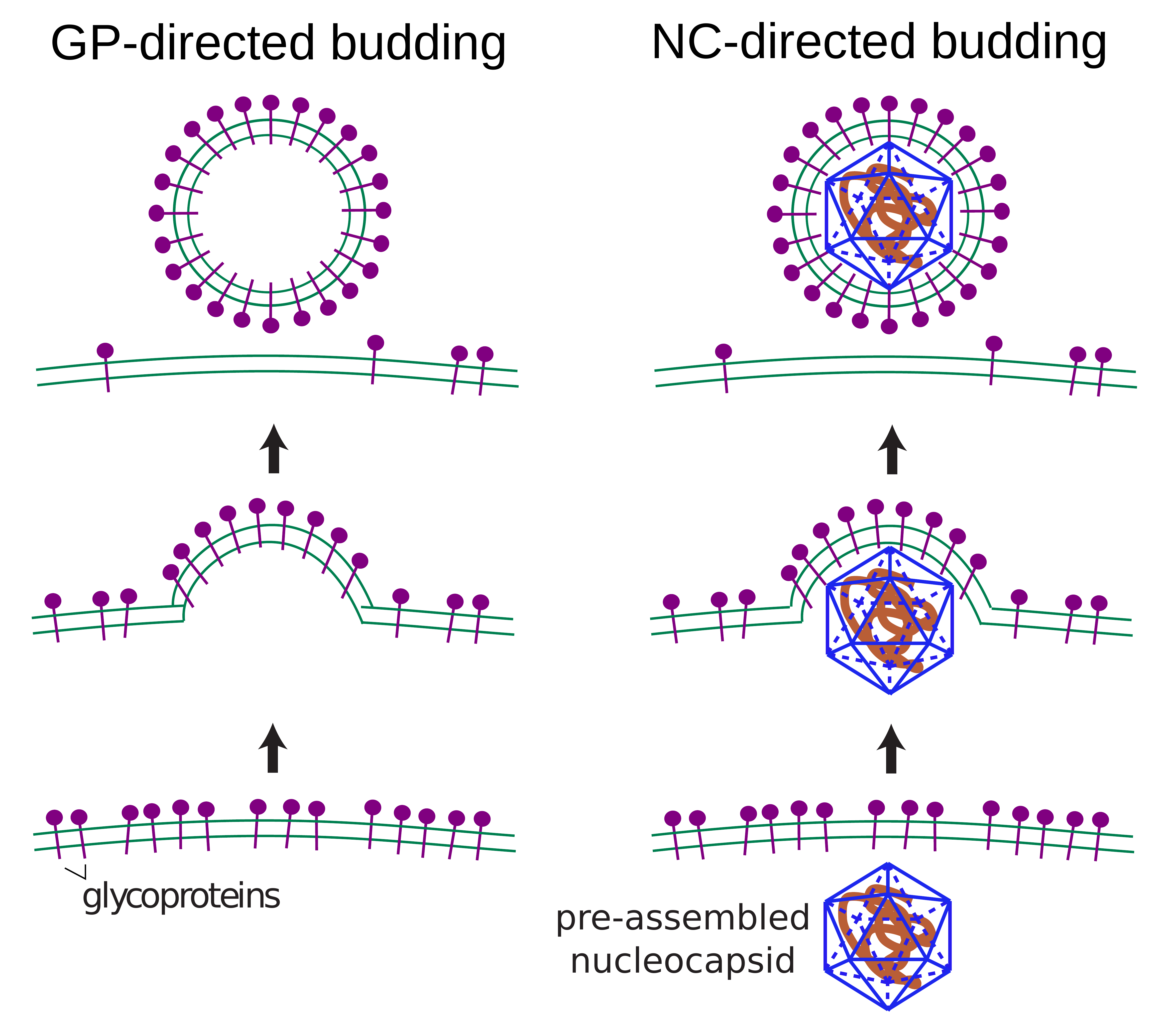}
    \caption{ Mechanisms of enveloped virus budding on membranes: glycoprotein(GP)-directed budding and nucleocapsid(NC)-directed budding. }
  \label{fig:budding}
  \end{center}
\end{figure}

In this article, we perform MD simulations on a coarse-grained model for GPs, the NC, and a lipid bilayer membrane to elucidate the forces driving enveloped virus budding. Our model is motivated by the alphavirus structure and experimental observations on alphavirus budding \cite{Forsell2000,Ruiz-Guillen2016,Garoff2004}, but we consider our results in the broader context of enveloped viruses. To evaluate the relative roles of a preassembled NC compared to the assembly of transmembrane glycoproteins in driving budding, we perform two sets of simulations. The first focuses entirely on glycoprotein-directed budding (Fig. \ref{fig:budding}a) by including only the membrane and model GPs, whose geometry and interactions drive formation of an icosahedral shell with the geometry of the alphavirus envelope. This model directly applies to experiments on budding from cells in which capsid assembly was eliminated  \cite{Forsell2000,Ruiz-Guillen2016}.  The second set of simulations includes model GPs and a preassembled NC, thus allowing for NC-directed budding (Fig. \ref{fig:budding}b).

We present phase diagrams describing how assembly morphologies depend on the strength of GP-GP and NC-GP interactions. The results demonstrate that the competition between the elastic energy of membrane deformations and deviations from preferred protein curvature can lead to polymorphic morphologies, and that templating by the NC can significantly decrease the resulting polydispersity. In the presence of a preassembled NC, there is a threshold strength of NC-GP interactions above which pathways transition from GP-directed to NC-directed budding. Our simulations enable visualization of the intermediates along each of these pathways as well as analysis of their relative timescales. In both pathways, assembly proceeds rapidly until budding is approximately 2/3 complete, after which curvature of the membrane at the neck of the bud imposes a barrier to subunit diffusion that significantly slows subsequent assembly and budding.  We discuss possible implications of this slowdown for enveloped viruses such as HIV that bud with incompletely formed capsids.

\begin{figure*}[hbt]
  \begin{center}
  \includegraphics[width=0.9\textwidth]{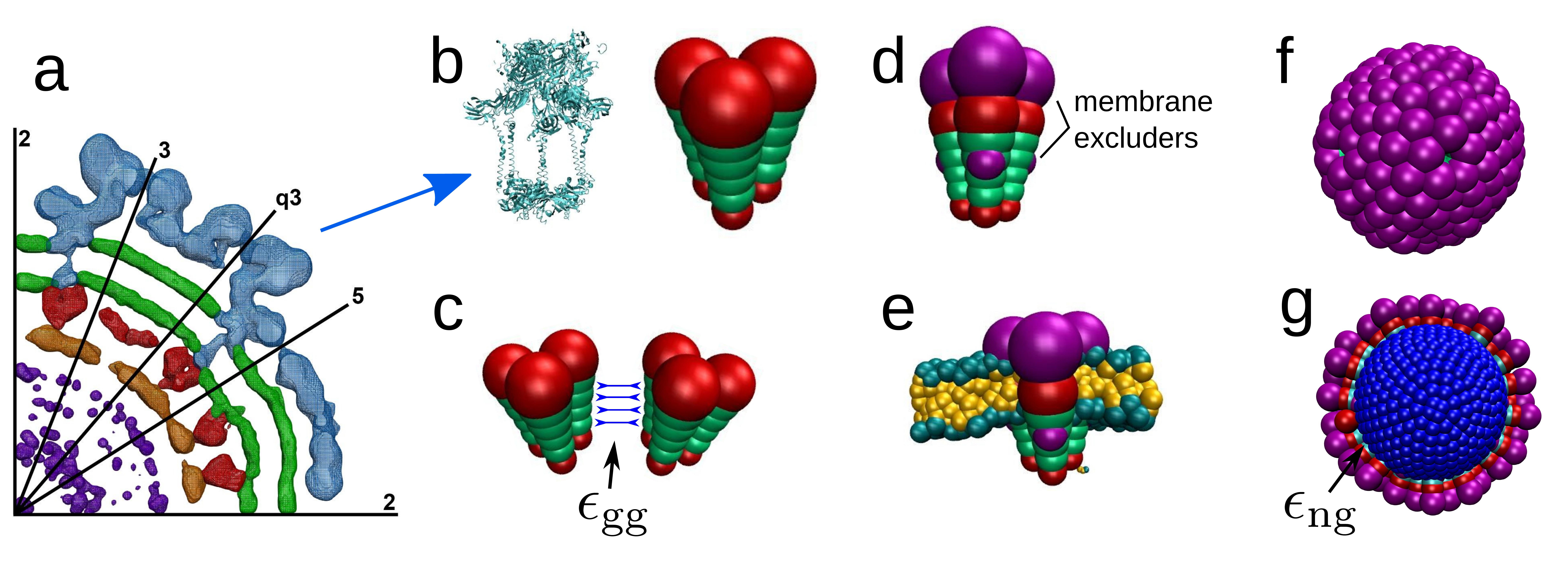}
    \caption{{\bf a)} Cryoelectron microscopy (CryoEM) density distribution of Sindbis virus. This central cross-section shows the inner structure of alphaviruses, with the RNA molecule (purple) enclosed by the NC (orange, red) and the lipid membrane (green) with the transmembrane GPs (blue). [{\bf b-g}] Computational model of the alphavirus GPs and NC. {\bf b)} Comparison between the GP trimer as revealed by CryoEM (PDB ID: 3J0C \cite{Zhang2002}) and our coarse-grained model trimer. Trimers are modeled as rigid bodies comprising three cones, with each cone formed by 6 pseudoatoms of increasing diameter. {\bf c)} Each of the 4 inner pseudoatoms of a cone (green) interacts with its counterpart in a neighboring cone through a Morse potential with well-depth $\egg$. All pseaudoatoms, including the `excluders' (red), interact through a repulsive Lennard-Jones potential.  {\bf d)} Model glycoproteins are trapped in the membrane by `membrane-excluders' (purple) which interact with membrane pseudoatoms through a repulsive Lennard-Jones potential.  {\bf e)} Complete trimer subunit embedded in the membrane. To aid visibility, in subsequent figures only the membrane excluders are shown. {\bf f)} Snapshot of a typical capsid assembled by model glycoproteins in the absence of a membrane, consisting of 80 trimer subunits. {\bf g)} Snapshot of typical capsid assembled by glycoproteins around the model nucleocapsid (NC, blue) in the absence of a membrane. The NC is modeled as a rigid spherical particle. NC pseaudoatoms interact with the lowermost pseudoatom in each GP cone through a Morse potential with depth $\eng$.}
  \label{fig:scheme}
  \end{center}
\end{figure*}

 \begin{figure*}[hbt]
  \begin{center}
  \includegraphics[width=0.8\textwidth]{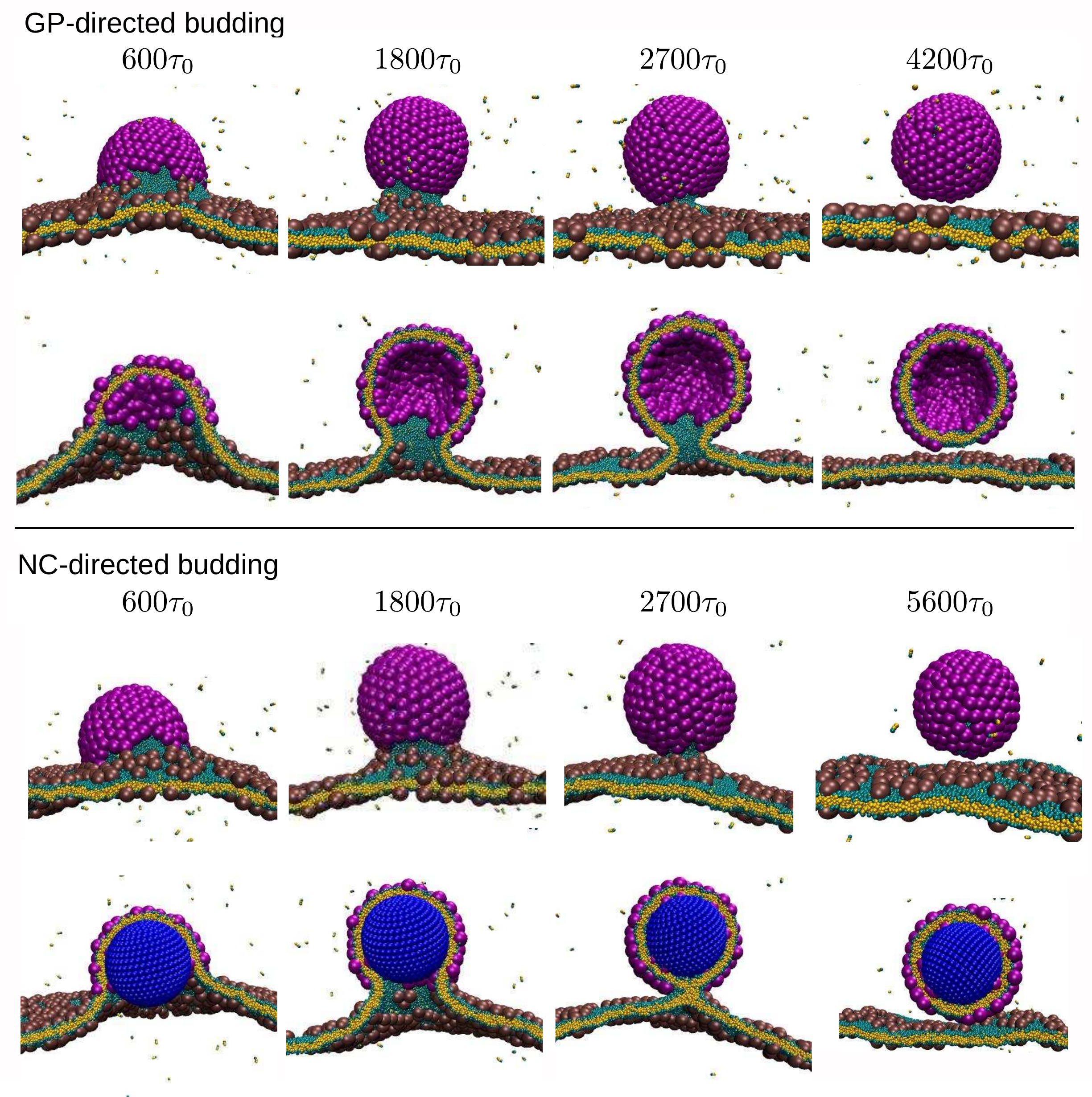}
    \caption{ {\it (Top)} Typical simulation trajectory of GP-directed budding, at simulation times 600, 1800, 2700 and 4200$\tau_{0}$, from left to right, with GP-GP interaction strength $\egg=2.3$.  {\it (Bottom)} Simulation trajectory of NC-directed budding, at 600, 1800, 2700 and 5600 $\tau_{0}$, with $\egg=2.5$ and NC-GP interaction strength $\eng=3.5$. The second timepoint in each row corresponds to an example of the intermediate with a constricted neck described in the text. Except where noted otherwise, the membrane bending modulus $\kmem\approx 14.5\kt$ throughout the manuscript.}
  \label{fig:trajectory}
  \end{center}
\end{figure*}

\begin{figure}
\caption{{\bf (A): Simulation Movie 1.} Animation of a typical simulation showing GP-directed budding, for $\egg=2.5$ and $\kmem=14.5 \kt$. Colors are as follows: GPs, magenta; membrane head groups, cyan; membrane tails, yellow. To show the membrane neck geometry more clearly the inactive subunits are rendered invisible in this animation. {\bf (B): Simulation Movie 2.} The same simulation trajectory as in {\bf (A)}, but rendered to show a central cross-section of the budding shell and the membrane. Inactive subunits are rendered brown in this animation. }
\label{fig:MovieGPdirected}
\end{figure}

\begin{figure}
\caption{{\bf (C): Simulation Movie 3.} Animation of a typical simulation showing NC-directed budding, for $\egg=2.5$, $\eng=3.5$, and $\kmem=14.5 \kt$. Colors are as in Simulation Movie 1, and the NC is colored blue. To show the membrane neck geometry more clearly we do not show inactive subunits in this animation. {\bf (D): Simulation Movie 4.} Animation showing a central cross-section of the NC-directed budding (same simulation trajectory as {\bf (C)}). Inactive subunits are shown in brown in this animation. }
\label{fig:MovieNCdirected}
\end{figure}


\section{Results}
\label{sec:results}
 Although intact viruses can be simulated at atomistic or near-atomistic resolution \cite{Freddolino2006,Zhao2013,Reddy2015,Perilla2015,Huber2016}, the time scales for alphavirus  assembly (ms-minutes) are prohibitive at such resolution. We thus consider a coarse-grained description for the viral GPs and the membrane, which enables tractable simulation of a large membrane over biologically relevant timescales while retaining the essential physical features of membranes and virus capsid and transmembrane proteins (see Fig. \ref{fig:scheme} and section \ref{sec:methods}). The membrane is represented by a solvent-free model which can be tuned to match properties of biological membranes while allowing simulation of large systems \cite{Cooke2005}.  Our model GPs are designed to roughly match the triangular shape, dimensions and aspect ratio of Sindbis virus GP trimers ~\cite{Mukhopadhyay2006,Zhang2002}. They experience lateral interactions, which in the absence of a membrane drive assembly into capsids containing 80 subunits, consistent with the 80 trimers in the alphavirus glycoprotein shell. In our simulations the GPs are embedded within the membrane, where they freely tilt and diffuse but cannot escape on simulation timescales.
  Motivated by the recent observation that alphavirus nucleocapsids do not require icosahedral symmetry \cite{Wang2015} to be infectious, we model the nucleocapsid as a rigid isotropic sphere. To account for experimental observations of capsid and GP conformational dynamics, the model GPs interconvert between assembly-inactive and assembly-active conformations which are respectively compatible or incompatible with assembly (see section~\ref{sec:methods}).

To compare GP-directed and NC-directed budding, we performed two sets of simulations that respectively included or did not include a NC. To understand how these pathways depend on parameters which can be controlled in experiments or varied under evolutionary pressures, we simulated assembly as a function of parameters controlling the GP-GP interaction strength $\egg$, the NC-GP interaction strength $\eng$ (when a NC is present), and the membrane bending modulus $\kmem$.  All energies are reported in units of the thermal energy, $\kt$. For notational convenience, we refer to particles assembled from GPs only as GP-particles, and GPs assembled around the NC as GPNC-particles.

\subsection*{Budding includes an intermediate with a constricted neck}
 We show snapshots from typical trajectories for simulations in the presence and absence of a NC in Fig. \ref{fig:trajectory}. In both cases, assembly and membrane deformation proceed rapidly until the GP shell is approximately 2/3 complete (the second timepoint in each row in Fig. \ref{fig:trajectory}), when the budding region is connected to the rest of the membrane by a narrow neck. Subunits within the neck experience restricted configurations due to the high membrane curvature. Thus, the neck acts as an entropic barrier that impedes subunit diffusion to the growing shell, causing the assembly rate to slow dramatically as the shell nears completion (Fig. \ref{fig:timescales}). The neck continues to narrow as additional GPs assemble until it becomes a tether connecting the bud and membrane.  In this article we do not consider ESCRT or related scission-inducing proteins, and thus the bud separates from the membrane only when a large thermal fluctuation induces membrane fission leading to scission of the tether.

\begin{figure}[hbt]
  \begin{center}
  \includegraphics[width=\columnwidth]{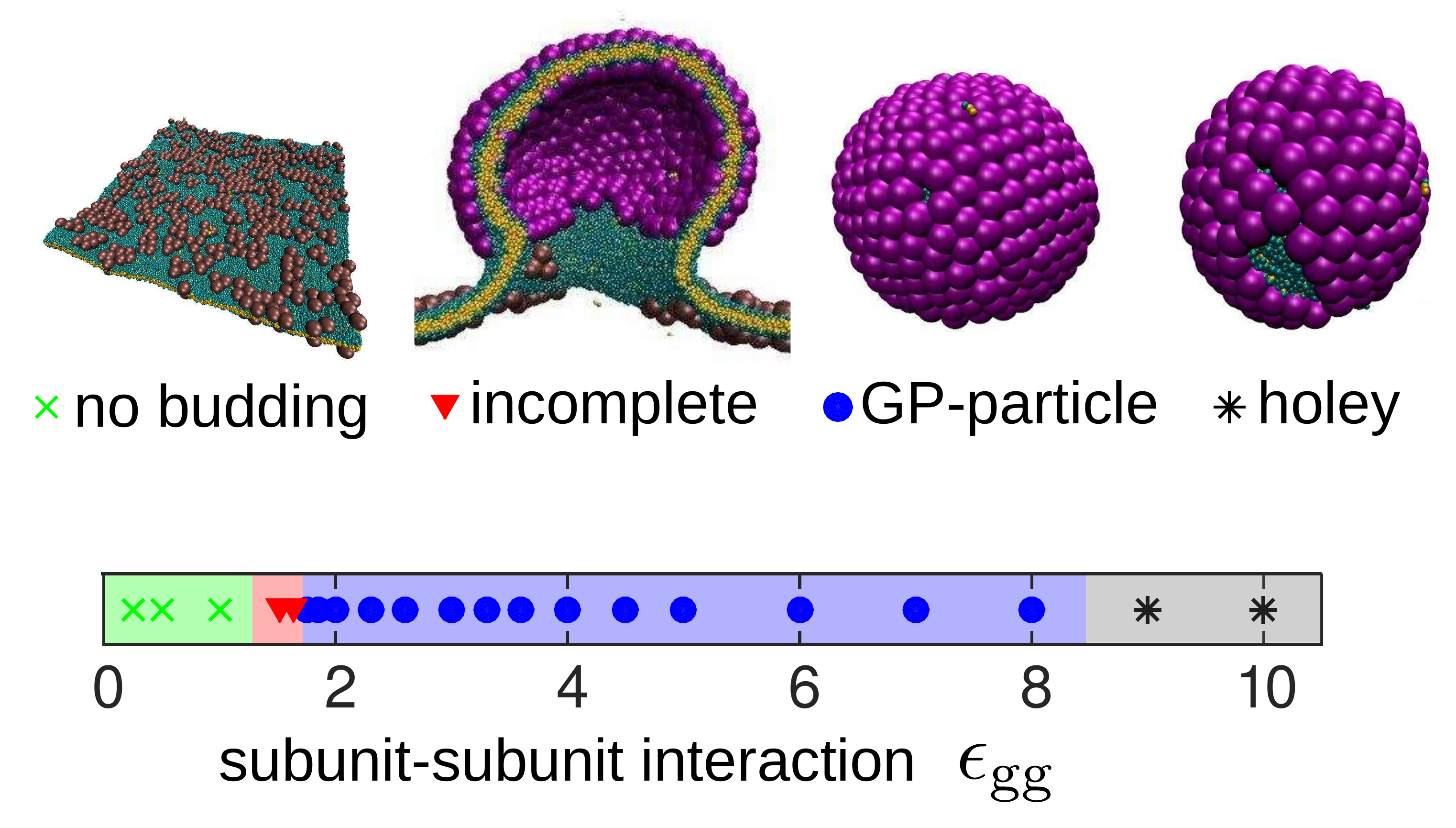}
    \caption{ Predominant end-products for assembly without a NC as a function of GP-GP interaction strength, along with simulation snapshots that exemplify each class of end-product.The distribution of end-products for several representative values of $\egg$ is shown in Fig.  \ref{fig:distributionoutcomes}.}
  \label{fig:phasediagramGP}
  \end{center}
\end{figure}

\subsection*{Glycoprotein-directed budding leads to complete but polydisperse particles}
\label{sec:gpdriven}
We first consider assembly in the absence of a NC, so that budding is necessarily GP-directed. 	Fig. \ref{fig:phasediagramGP} shows the most frequent end-product obtained as a function of the GP-GP interaction strength. For weak interactions ($\egg<1.4$) assembly is unfavorable.  In contrast, bulk simulations (model GPs in the absence of a membrane) exhibit shell assembly for $\egg>0.97$ at trimer concentration $\phi_{3D}=3.4 \cdot 10^{-5}\sigma^{-3}$ (see Fig. ~\ref{fig:capsidsbulk}a in the SI), well below the effective concentration of trimers on the membrane $\phi_{2D}^{eff}=0.001\sigma^{-3}$ \footnote{To compare the subunit concentration in bulk $\phi_{3D}=N_{s}/L^{3}$, where $N_{s}$ is the number of subunits and $L$ the box size, with the subunit concentration on the membrane, we measure the mean deviation of the height fluctuations of the subunits on the membrane, $\lambda\approx 4.5$, and the effective concentration on the membrane is then given by  $\phi_{2D}^{eff}=N_{s} / L^2 \lambda$.} . This result demonstrates that the membrane rigidity can introduce a substantial barrier to assembly~\cite{Ruiz-Herrero2012}.

Within a narrow range of interaction strengths  $1.4< \egg<1.7$, assembly and budding stalls at the constricted neck intermediate described above. For these parameters, the intermediate remains upon extending the simulation length to 10,500$\tau_{0}$, suggesting that it corresponds to a true steady state or a very long-lived kinetic trap.  This configuration resembles partially assembled states that were predicted theoretically \cite{Zhang2008,Foret2014}, but arises due to different physics. We find that the range of $\egg$ over which the state arises depends on the subunit geometry, but the state exists for any geometry we considered (see section \ref{sec:subunit} in the SI). A similar configuration was observed during simulations of assembly and budding of a 12-subunit capsid on a membrane \cite{Ruiz-Herrero2015}, suggesting it is a generic feature of assembly and budding. However, in that work assembly never proceeded past the partially assembled state for any parameter set on a homogeneous membrane, possibly due to the to the small size of the simulated capsid.

For stronger interactions, we observe complete budding. However, the morphologies of the resulting GP-particles depend on the interaction strength in two ways. First, overly strong interactions ($\egg>9.0$) drive rapid assembly which can proceed simultaneously along multiple fronts within a shell, leading to the formation of holey GP-particles (Fig. \ref{fig:phasediagramGP}). This result is consistent with holey capsids that assemble under strong interactions in bulk simulations \cite{Hagan2006,Rapaport2012}. Second, for moderate interactions ($1.6< \egg<9.0$) the shells are complete, but their size  depends on the interaction strength, with  typical sizes ranging over 95--140 subunits. The origin of this polymorphism is  discussed later in this section.

\subsection*{Nucleocapsid-directed budding leads to more monodisperse particles}
\label{sec:ncdriven}

\begin{figure*}[hbt]
  \begin{center}
  \includegraphics[width=0.8\textwidth]{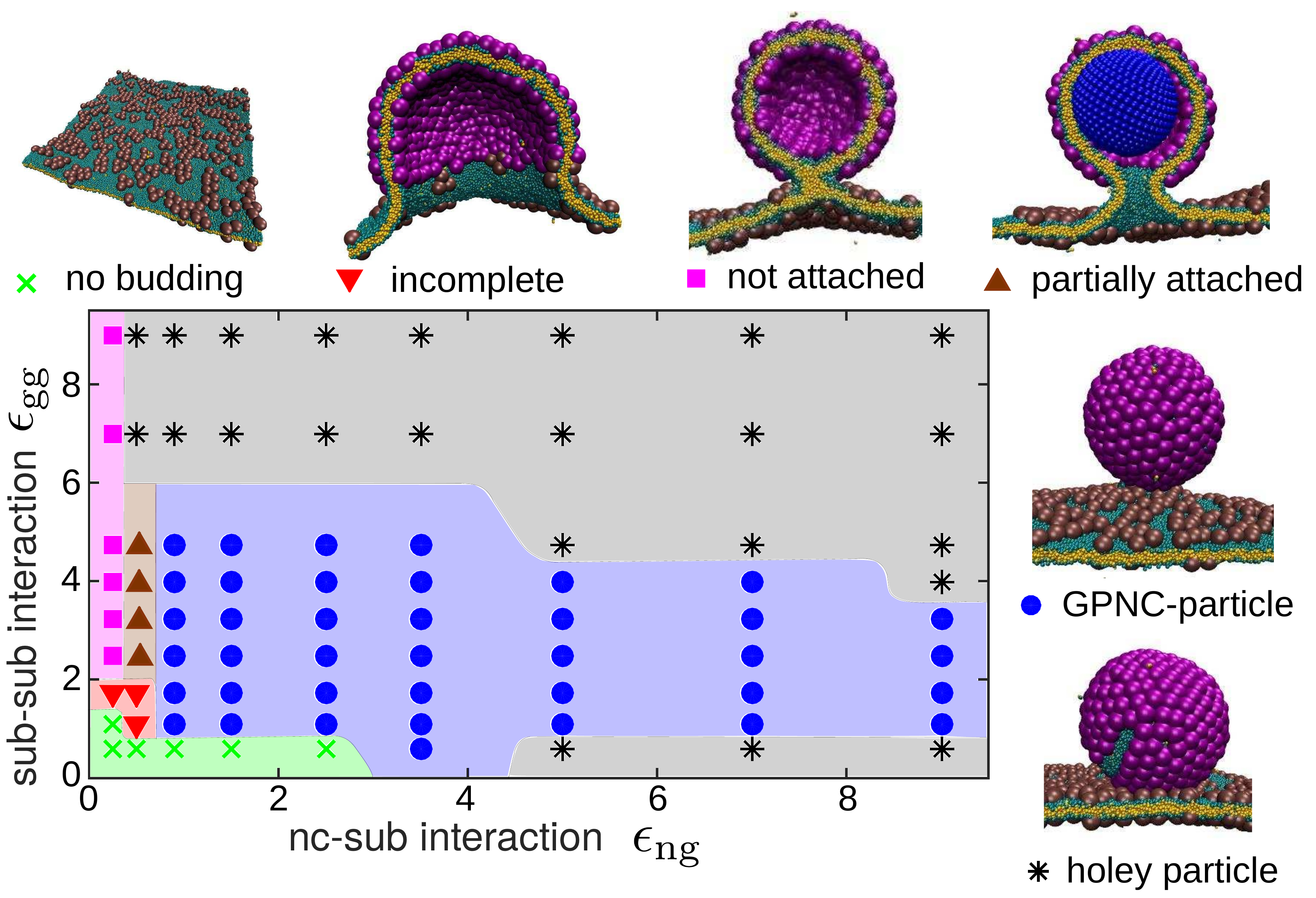}
    \caption{ Predominant end-products of the NC-directed budding, as a funtion of the NC-subunit interaction $\eng$ and the subunit-subunit interaction $\egg$, and snapshots showing representative examples of each outcome. }
  \label{fig:phasediagramNC}
  \end{center}
\end{figure*}

The predominant end-products of assembly in the presence of a NC are shown in Fig.~\ref{fig:phasediagramNC} as a function of the two interaction parameters: GP-GP ($\egg$) and NC-GP ($\eng$).
We observe complete assembly  and budding  for $\eng > 0.9$ and $1\le \egg \le 6$ (Fig.~\ref{fig:phasediagramNC}, blue region), low to moderate GP-GP interactions. Compared to the GP-directed pathway, the presence of a NC allows assembly to occur at a lower $\egg$ as evidenced by obtaining complete shells even for $\egg$ as low as 0.9.
Outside of this range, several other end-products arise. For $\eng<0.4$ (Fig.~\ref{fig:phasediagramNC}, pink region), the NC interactions are sufficiently weak that budding is entirely GP-directed (\ie GP shells assemble and bud, but not around the NC). In the range  $0.4< \eng<0.9$ (Fig.~\ref{fig:phasediagramNC}, brown region) we observe an intermediate regime in which the NC promotes nucleation but fails to act as a perfect template. The GP shell initially starts assembling on the NC surface, but eventually separates from the surface to form a larger shell due to the effect of membrane rigidity (see Section \ref{sec:polymorphism}). The result is an asymmetric shell that is partially attached to the NC; with a typical size of 95 subunits it is smaller than a GP-particle but considerably larger than the intrinsic preferred shell size. These two assembly outcomes (GP shells partially attached or unattached to the NC) demonstrate that the presence of a NC does not necessarily imply NC-directed budding; there is a minimum NC-GP interaction strength required for the NC to direct the assembly and budding pathway.

Strong GP-GP interactions ($\egg>6$, Fig.~\ref{fig:phasediagramNC}, upper grey region) lead to holey particles. This result can be explained as in the case of holey GP-particles described above; however, notice that the threshold value of $\egg$ is smaller than in the absence of the NC ($\egg=9$).  Interestingly, we also observe holey GPNC-particles when strong NC-GP interactions are combined with weak GP-GP interactions ($\egg<1.0$, Fig.~\ref{fig:phasediagramNC}, lower grey region). In this regime, NC uptake proceeds rapidly, but GP subunits do not associate quickly enough form a complete shell as budding proceeds.  

To further elucidate the interplay between the two interactions, Fig.~\ref{fig:energy} (SI) compares the total energetic contributions from GP-GP and NC-GP interactions for budded GPNC-particles.  We see that GP-GP interactions account for the majority of the attractive energy stabilizing the shell, with the NC-GP providing as little as 10-20\% of the total energy. These results highlight the delicate balance between GP-GP and NC-GP interactions required to obtain a well-formed GPNC-particle.

\subsection*{Membrane-induced polymorphism}
\label{sec:polymorphism}

 Although both GP-only budding and NC-directed budding lead to the formation of complete particles, the morphology of budded shells significantly differs between both mechanisms. Fig. \ref{fig:bending}a shows the mean shell size as a function of interaction strength (averaged over all closed particles). For GP-directed budding, we see a strong dependence of particle size on subunit interactions: weak interactions lead to ovoid particles containing up to 140 trimers. As $\egg$ increases, particles become smaller and more spherical, more closely resembling the shells that assemble in bulk simulations. We show snapshots of typical GP-shells assembled at weak and strong interactions. On the contrary, the size of GPNC-particles is nearly constant with $\egg$ and only slightly larger (81-83 subunits) than the preferred size in bulk simulations.

{\textbf{A theoretical model for membrane-induced polymorphism.} Although shell assembly is necessarily out-of-equilibrium in finite-length simulations, we can understand the dependence of size on interaction strength from a simple equilibrium model that uses the Helfrich model \cite{Helfrich1973} to account for the elastic energy associated with membrane deformation (which has no spontaneous curvature and thus favors flat configurations) and deviation of the GP shell from its preferred curvature. The calculation is detailed in the Appendix. Minimizing the total free energy for a system with fixed number of GP subunits obtains that the most probable number of subunits in a shell $n$ corresponds to the value which minimizes the elastic energy per subunit, given by
\begin{align}
n = n_{0} \left( 1 + \frac{\kmem}{ \kcap} \right)^{2},
\label{eq:nsubs}
\end{align}
where $n_{0}$ is the number of subunits in the equilibrium configuration in the absence of a membrane ($n_{0}=80$ in our model), and  $\kmem$ and $\kcap$ as the membrane and shell bending moduli. Thus the preferred GP-particle size is determined by the ratio $\kmem / \kcap$, which quantifies the competition between the membrane and shell deformation energies.
 Only in the limit where the shell rigidity dominates, $\kmem / \kcap \to 0$, will GP-particles exhibit the size observed in bulk simulations.

To compare the theoretical estimate to the shell sizes observed in simulations, we estimated the relationship between the GP interaction strength and the shell bending modulus as $\kcap \approx 25.66 \egg$ (we show the complete derivation in section \ref{sec:bendingestimation} in the SI), leading to a range of shell bending rigidities of $\approx 40-250\kt$.  This range coincides with bending rigidity values measured in AFM experiments on virus capsids (see the Discussion) \cite{Michel2006,Roos2007,May2011}. The prediction of Eq.~\eqref{eq:nsubs} using this estimate and the estimated membrane rigidity $\kmem=14.5\kt$ is shown in Fig \ref{fig:bending}a.  The prediction is also compared against simulated particle sizes as a function of the parameter $\kmem / \kcap$ for different membrane bending rigidities in Fig.~\ref{fig:bending}b.  For moderate values of $\kmem / \kcap$  we observe good agreement between the theory and simulation results, especially considering that there is no fit parameter. The agreement breaks down for $\kmem / \kcap\gtrsim0.3$, likely for several reasons. Firstly, our theory assumes a closed GP shell, whereas the size of the incomplete region of the GP shell increases in size with $\kmem / \kcap$, as illustrated by the snapshots in Fig.~\ref{fig:bending}b. Secondly, subunits within the largest GP particles are far from their preferred interaction angle, and thus their elastic response could be nonlinear. Finally, finite-size effects could become non-negligible for the largest buds.

In contrast to GP-directed assembly, Figs.~\ref{fig:phasediagramGP} and \ref{fig:phasediagramNC} demonstrate that a NC can dramatically change the morphology of a GP shell by acting as a template over a broad range of interaction strengths. The observed monodispersity in GPNC-particles can be understood from Eq.~\eqref{eq:nsubs} by noting that the NC is modeled as a perfectly rigid sphere in our simulations, and thus corresponds to the limit $\kcap\rightarrow \infty$ if it acts as a perfect template for the GP shell. The relevance of this approximation to enveloped viruses is considered in the Discussion.

\begin{figure}[hbt]
  \begin{center}
  \includegraphics[width=\columnwidth]{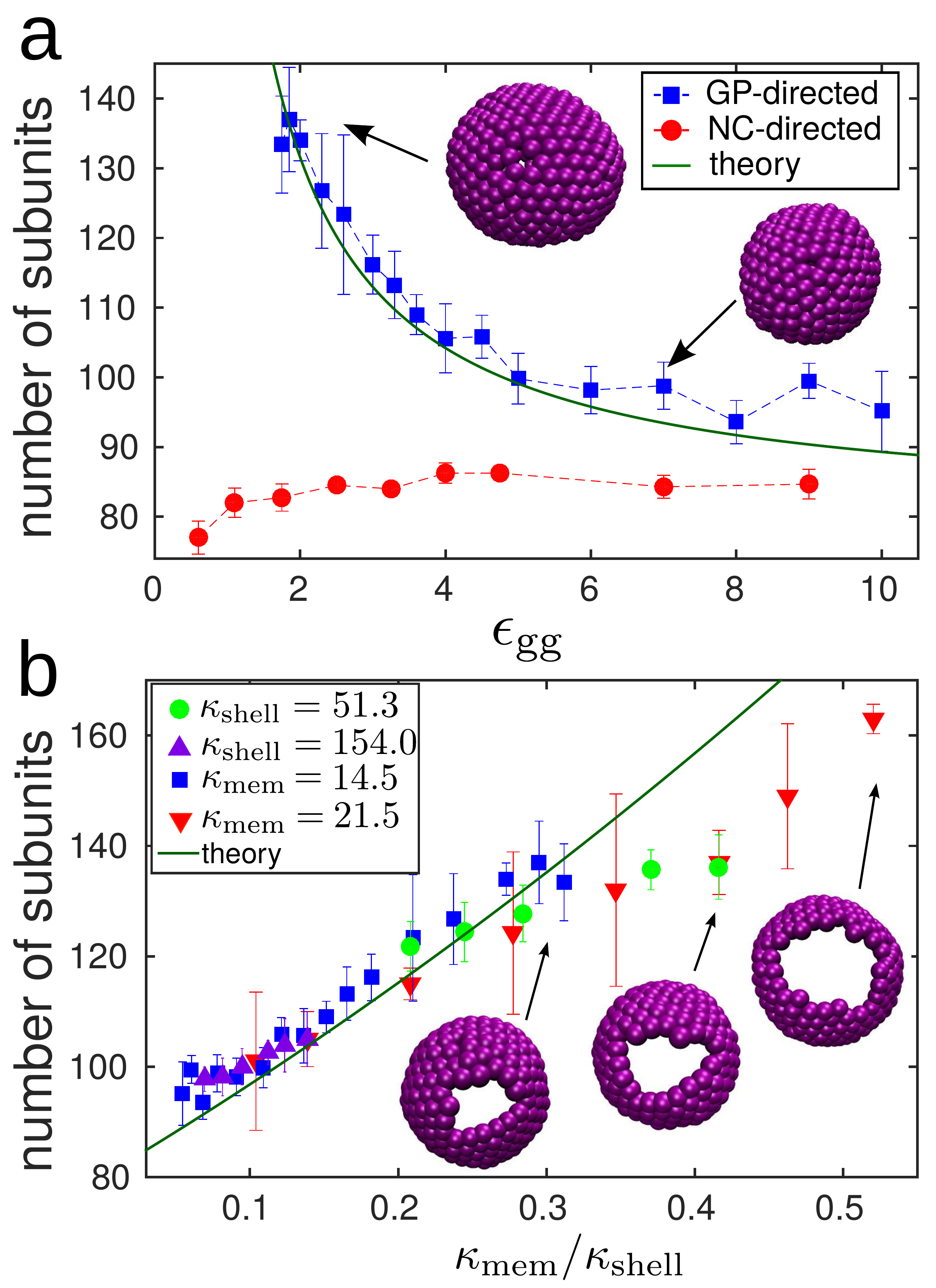}
    \caption{ {\bf a)} Average number of subunits in budded particles as a function of GP-GP affinity, for  GP-directed budding (\textcolor{blue}{$\blacksquare$} symbols) and NC-directed budding with $\eng=4.5\kt$ (\textcolor{red}{$\CIRCLE$} symbols). The solid green line gives the theoretical prediction (Eq.~\eqref{eq:nsubs})  for the estimated capsid rigidity $\kcap=25.66 \egg$ (see section \ref{sec:bendingestimation} (SI)) and $\kmem=14.5\kt$. {\bf b)} Average number of subunits in GP-shells as a function of the ratio between membrane and shell bending modulii, $\kmem / \kcap$. The data includes different sets of simulations in which either $\kmem $ or $\kcap$ is maintained constant, and we sweep over the other parameter: $\kmem=14.5$ (\textcolor{blue}{$\blacksquare$}), $\kmem=21.5$ (\textcolor{red}{$\blacktriangledown$}), $\kcap=51.3$ (\textcolor{green}{$\CIRCLE$}), and $\kcap=154.0$ (\textcolor{dpurple}{$\blacktriangle$}). The theoretical prediction (Eq.~\eqref{eq:nsubs}) is shown as a black solid line.}
  \label{fig:bending}
  \end{center}
\end{figure}

\subsection*{The nucleocapsid influences timescales for late stage budding}
\label{sec:timescales}
As noted above assembly can be divided into two stages. The shell grows rapidly until about 2/3 completion, after which neck curvature significantly slows subunit association (Fig. \ref{fig:timescales}). The timescale for the second stage depends on the interaction parameters and whether a NC is present --- for the small GPNC-particles assembly is completed quickly $\sim 450\tau_{0}$, whilst in the large GP-particles with the broadest necks it may require up to $3,000\tau_{0}$. In contrast, the timescale for the first stage is almost independent of interaction strengths and the NC, and depends only weakly on membrane bending modulus.  Furthermore, as shown in section \ref{sec:conformational} (SI), conformational switching is not rate limiting, implying that assembly rates during the first stage are limited by subunit diffusion.

This observation parallels models for clathrin-independent receptor-mediated endocytosis, in which the endocytosis timescale is estimated from the time required for membrane receptors to diffuse to the enveloped particle \cite{Gao2005,Bao2005}.  Applying the same analysis to our simulations, the timescale for GPs to diffuse to the budding site is given by $\tau\sim l^{2}/2D$, with $D_{\text{sub}}=\sigma^{2}/\tau_{0}$ the GP diffusion constant in our simulations, and $l\approx 45\sigma$ as the radius of the region around the budding site initially containing 80 trimers, enough to envelop the particle. This estimate yields $\sim1200\tau_{0}$, which is reasonably close to the typical timescale for stage 1 observed in the simulations, $\sim1000\tau_{0}$. Note that this model does not describe the timescale of the latter stage of assembly, since the curved neck region imposes a barrier to subunit diffusion which increases as the particle nears completion.

\subsection*{GP conformational changes avoid kinetic traps}
Finally, we note that when subunit conformation changes are not accounted for (\ie all subunits are in the active state), we observe complete assembly and budding only under a narrow range of interaction strengths for NC-directed budding and not at all for GP-directed budding (section \ref{sec:conformational}, SI).  For most interaction strengths the simulated densities of GPs led to multiple small aggregates which failed to drive significant membrane deformation. This behavior is indicative of kinetic trapping, known to occur in assembly reactions at high concentrations or binding affinities \cite{Hagan2014,Whitelam2015}. The ability of an inactive conformation to avoid this trap is consistent with simulations of bulk assembly \cite{Lazaro2016,Grime2016}, and the ability of budding to proceed in the presence of high subunit concentrations (when conformational changes are accounted for) is consistent with the observation of high densities of GPs in the membranes of cells infected with Sindbis virus \cite{Bonsdorff1978}.

\begin{figure}[hbt]
  \begin{center}
  \includegraphics[width=\columnwidth]{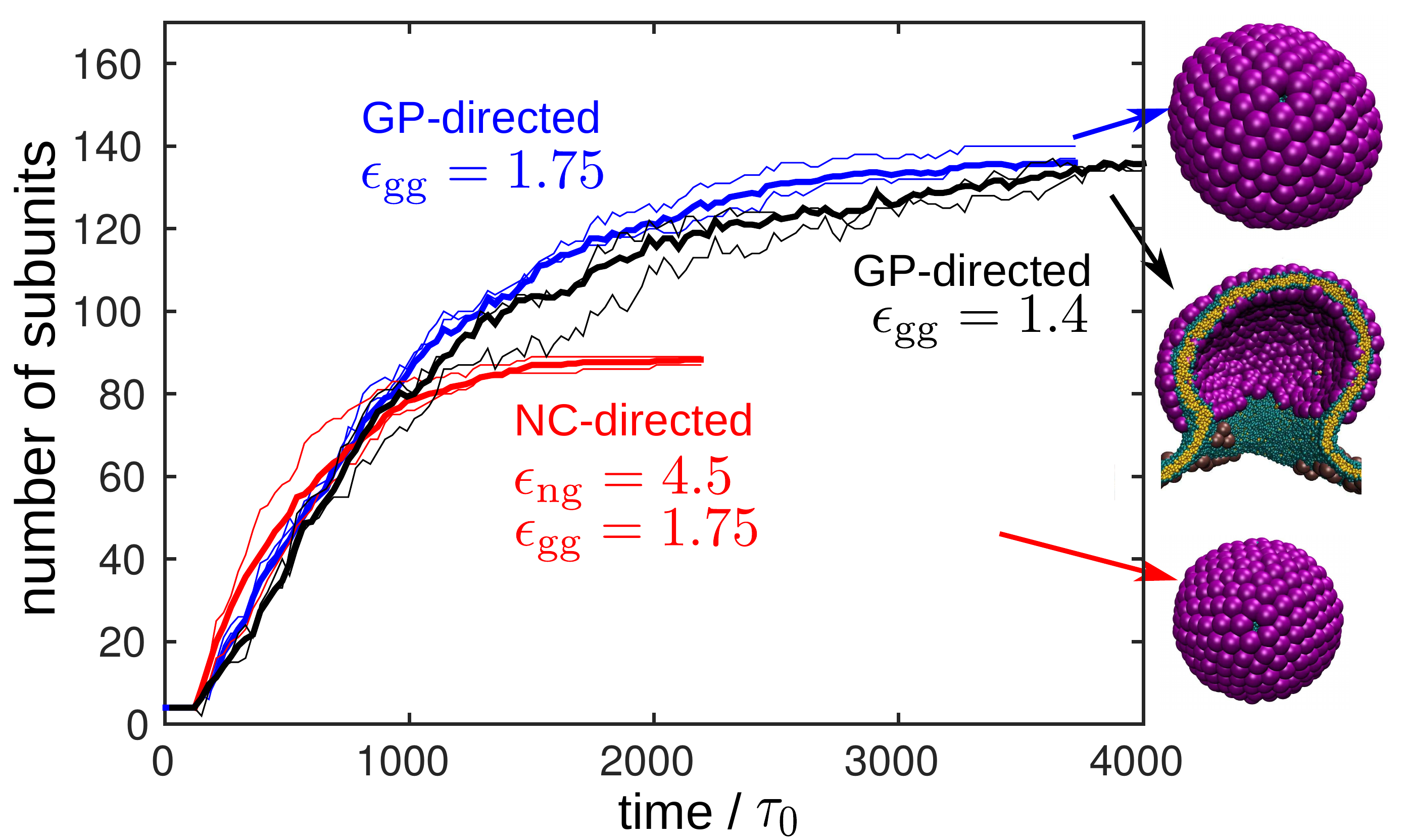}
    \caption{ Number of GP trimers in budding shells as a function of time for trajectories at different parameter values: GP-directed budding with $\egg=1.4$ (black) or $\egg=1.75$ (blue) and NC-directed budding with  $\eng=4.5$ and   $\egg=1.75$ (red). The snapshots to the right show the assembly products. For each parameter set the thick line shows an average over three trajectories, and the two thin lines show two individual trajectories to give a sense of the size of fluctuations. The lines end when budding occurs, except for the GP-directed case with $\egg=1.4$ which ended in the stalled partially budded state.  }
  \label{fig:timescales}
  \end{center}
\end{figure}


\section{Discussion and conclusions }
\label{sec:discussion}
\subsection*{Advantages of NC-driven budding} 
We have described dynamical simulations of the assembly and budding of GPs in the presence and absence of a preassembled NC, and presented phase diagrams showing how assembly pathways and products depend on the relevant interaction parameters.
The key difference between NC-directed and GP-directed budding identified by our results  is variability of the budded particle size and morphology.
The presence of the core directs the morphology of the particle, which  may have direct consequences on particle stability during transmission and the conformational changes that occur during particle entry into a new host cell.
The number of GPs in GP-particles (containing no NC)  varies by $>$50\% over the range of interaction parameters in which we observe successful budding, in comparison to a variation of less than 5\% for GPNC-particles (containing a NC). A simple equilibrium model accounting for the competition between the energy costs associated with membrane deformation and deviation of the GP shell from its preferred curvature was qualitatively consistent with the simulation results. The membrane bending energy favors formation of larger particles, while a higher bending rigidity of the GP shell favors smaller particles.  This prediction applies to any form of assembly on a fluid membrane, and thus is relevant to capsid-directed budding as well as the GP-directed budding studied here.

It is worth considering this observation in the context of recent experimental observations on budding in the absence of a NC, as well as experimental measurements on viral particle elastic properties.
Several experimental studies have reported GP-directed budding from cells in which the NC proteins are impaired \cite{Forsell2000,Ruiz-Guillen2016}. In particular, Ruiz-Guillen \etal \cite{Ruiz-Guillen2016} recently showed that cells expressing the genome and GPs, but not the capsid protein, for Sindbis and Semliki Forest virus generate infectious viral particles that can propagate in mammalian cells.  For the wild-type virus, in which the GPs assemble around the NC, they estimated a virion diameter of 60-70nm, whereas the GP-only particles were typically 100-150nm. This increase in particle size is consistent with our simulations and theoretical model, suggesting that the relatively low rigidity of the GP shell leads to the formation of large particles.  However there are two important caveats to this interpretation.  First, the bending rigidity of the alphavirus GP shell has not been measured, so we cannot directly predict the increase in particle size.  Second, our model assumes that the preferred curvature of the GP shell is commensurate with its size in a wild type virion, for which there is no direct evidence. A recent study of herpes simplex virus nuclear egress complex (NEC), which consists of two viral envelope proteins, found that NEC particles budded in the absence of capsid are smaller than native viral particles \cite{Hagen2015}. In our model this observation would require that the intrinsic spontaneous curvature radius of the NEC complex is smaller than that of the capsid, as suggested by the authors \cite{Hagen2015}.

Despite different tendencies for polymorphism, our results show that GP-directed and NC-directed budding share many similarities, presenting a challenge for distinguishing between NC-directed and assembly-directed budding.
For example, assembly timescales depend only weakly on the presence of a NC and interaction parameters, instead being limited by diffusion of GP subunits to the budding site. We anticipate that this result would be unchanged by adding additional ingredients to the model, since the conditions we simulated (high density of GPs in the surrounding membrane) correspond to the lower bound of GP diffusion timescales.  The same principle would apply if GPs are directly targeted to the budding site (rather than diffusing on the membrane) as has been suggested for alphavirus budding \cite{Martinez2014}, except that the diffusive flux would be replaced by a targeting flux.

\subsection*{Requirement for scission machinery}
A second commonality between the two scenarios investigated is that assembly slows down considerably after the GP shell reaches approximately 2/3 completion, because the high curvature of the neck region imposes a barrier to subunit diffusion. For weak interactions, this leads to a long-lived partially budded state. For stronger interactions budding eventually completes, but completion of the shell can be preempted by scission.
 Since spontaneous scission is a rare event, most viruses actively drive scission by either recruiting host cell machinery, such as the ESCRT protein complex in the case of HIV \cite{Baumgartel2011}, or encoding their own macinery, such as the M2 protein in Influenza \cite{Rossman2010}. In alphaviruses specifically the scission machinery has not been identified, though it is known that alphavirus budding is independent of ESCRT proteins \cite{Chen2008}. Although in the present article we do not consider the action of these scission-inducing mechanisms, our observation of a slowdown of assembly rates due to neck curvature could be relevant to HIV budding. If the association rate becomes sufficiently slow, ESCRT-directed scission will occur before assembly completes, leaving up to 40\% of the shell incomplete, as is observed for immature HIV virions \cite{Briggs2009}. In support of this possibility, we note that although scission is a rare event in our simulations because there is no ESCRT, it usually occurs before the final 1-3 subunits assemble causing the budded particle to have small hole at the scission site.




\subsection*{Virus elasticity and GP spontaneous curvature affect budding morphology}
Since Fig. \ref{fig:phasediagramGP} suggests that the elastic properties of the different viral components are key determinants of the assembly product,
it is worth considering the validity of the model parameters.
The mechanical properties of viruses have been extensively studied using atomic force microscopy (AFM) indentation \cite{Michel2006,Roos2007} and fluctuation spectrum analysis \cite{May2011}. Typical estimates lie within the range $\kappa=30-400\kt$, with considerable variation depending on the specific virus and experimental technique. A recent work by \citet{Schaap2012} explicitly compared the stiffness of the capsid protein coat of influenza virus with that of a lipid membrane, by AFM indentation of similarly sized particles, with the goal of identifying the contribution of the matrix proteins to the virus stiffness. They found that matrix coats are approximately ten times stiffer than bare membranes,  $\kmem / \kcap \sim 0.1$. This value lies within the range explored in our simulations.
Similarly,  Kol \etal \cite{Kol2007} investigated the effect of HIV maturation on its mechanical properties.
The immature HIV particle consists of a gag polyprotein capsid surrounded by a lipid bilayer containing viral envelope proteins. During maturation, the NC and capsid portions of gag are cleaved, leaving only a thin matrix layer and the envelope proteins in contact with the bilayer. Kol \etal found that this cleavage softens the particles by an order of magnitude, suggesting that inter-protein contacts of the underlying capsid layer are necessary for the high rigidity of immature virions.

Finally, we note that the membrane elasticity properties also play a role in determining particle morphology. Depending on the virus family and host cell type, enveloped viral particles bud through different cellular membranes (the plasma membrane, the ER, the ERGIC, or the nuclear membrane), all of which have different lipid compositions and thus different bending properties. Moreover, many viruses create and/or exploit membrane microdomains with different compositions (such as lipid rafts) as preferential locations for budding \cite{Waheed2010,Welsch2007,Rossman2011}. The effect of inhomogeneous membrane elastic properties on particle morphology thus deserves further exploration. While these ingredients can be incorporated into the model, the results described here demonstrate that the interplay between the elastic properties of membranes and viral proteins and the presence of an interior core can shape the morphology of a budding particle.


\section{Methods}
\label{sec:methods}

\subsection{Glycoproteins and capsid}
\label{sec:capsid}
Our coarse-grained GP model is motivated by the geometry of Sindbis virions as revealed by cryoelectron microscopy ~\cite{Mukhopadhyay2006,Zhang2002}. The outer layer of Sindbis is comprised from heterodimers of the E1 and E2 GPs.  Three such heterodimers form a tightly interwoven trimer-of-heterodimers, and 80 of these trimers are organized into a T=4 lattice. On the capsid surface each trimer forms a roughly equilateral triangle with edge-length $\sim 8$nm. In the radial direction, each E1-E2 heterodimer spans the entire lipid membrane and the ectodomain spike, totaling $\sim12$nm  in length. In our model, we consider the GP trimer as the basic assembly subunit, assuming that the formation of trimers is fast relative to the timescale for assembly of trimers into a complete capsid.

Our subunit model aims to capture the triangular shape, aspect ratio, and preferred curvature of the GP trimers while minimizing computational detail. To this end, we employ the conical particles studied by Chen \etal \cite{Chen2007}, modified in two ways.
First, our GP trimer subunit comprises three cones, which are fused together and simulated as a rigid body. Second, the cones are truncated, so that they form a shell with an empty interior, as shown in Fig. \ref{fig:scheme}. The cone length and trimer organization within the capsid are consistent with the Sindbis structure (see section \ref{sec:subunit} (SI) for full details). Note that while the domains primarily responsible for curvature of alphavirus GPs are located to the exterior of the envelope, the conical regions which drive curvature of the model subunit oligomers are located within and below the plane of the membrane. We found that this arrangement facilitated completion of assembly (see section \ref{sec:subunit} in the SI).

Each cone consists of a linear array of six beads of increasing diameter.
Two nearby cones experience repulsive  interactions,
 mediated by a repulsive Lennard-Jones potential between all pairs of beads, with size parameter $\sigma$ equal to the bead diameter.  In addition, each of the four inner beads experiences an attractive interaction with its counterpart (the bead with the same diameter) in the neighboring subunit, modeled by a Morse potential. The Morse potential depth $\egg$ determines the subunit-subunit interaction strength, which is related to the GP-GP binding affinity. The equilibrium distance of the Morse potential $\re$, and the Lennard-Jones diameter $\sigma$ for each interacting pair is chosen to drive binding towards a preferred trimer-trimer angle.  We set the
preferred angle so that in bulk simulations (in the absence of membrane) the subunits predominantly assemble into aggregates with the target size, 80 subunits. However, there is a small amount of polydispersity, with some capsids having sizes between 79 and 82 subunits (Fig.~\ref{fig:capsidsbulk}). Although the subunit geometry locally favors hexagonal packing, formation of a closed capsid requires 12 five-fold defects \cite{Grason2016}. We find that the spatial distribution of these defects is typically not fully consistent with icosahedral symmetry for dynamically formed capsids. It is unclear whether this is a kinetic effect or indicates that icosahedral symmetry is not the free energy minimum at this particle size. However, the relatively high monodispersity observed suggests that the 80-subunits capsid is a free energy minimum and assembly is robust at these conditions.

\subsection{Lipid membrane}
\label{sec:membrane}
The lipid membrane is represented by the implicit solvent model from Cooke and Deserno \cite{Cooke2005}. This model enables on computationally accessible timescales the formation and reshaping of bilayers with physical properties such as rigidity, fluidity, and diffusivity that can be tuned across the range of biologically relevant values. Each lipid is modeled by a linear polymer of three beads connected by FENE bonds; one bead accounts for the lipid head and two beads for the lipid tail.  An attractive potential between the tail beads represents the hydrophobic forces that drive lipid self-assembly. In section \ref{sec:bendingestimation} in the SI we estimate the bending rigidity of the membrane in our simulations by analyzing their fluctuations spectra. Unless otherwise specified, our simulations used $\kmem\approx 14.5 \kt$ as a typical rigidity of plasma membranes.

\subsection{Glycoprotein-membrane interactions}
The effect of individual GPs on the behavior of the surrounding membrane has not been well characterized. Moreover, to facilitate interpretation of our simulation results, we require a model in which we could independently vary subunit-subunit interactions and subunit-membrane interactions. Therefore, we use the following minimal model for the GP-membrane interaction.
We add six membrane excluder beads to our subunit, three at the top and three at the bottom of the subunit, with top and bottom beads separated by 7nm (Fig.~\ref{fig:scheme}c,d).  These excluder beads interact through a repulsive Lennard-Jones potential with all membrane beads, whereas all the other cone beads do not interact with the membrane pseudoatoms. In a simulation, the subunits are initialized with membrane located between the top and bottom layer of excluders.  The excluded volume interactions thus trap the subunits in the membrane throughout the length of the simulation, but allow them to tilt and diffuse laterally. Separating the subunit pseudoatoms that interact with the membrane from those which control the subunit-subunit potential allows us to independently vary subunit-subunit and subunit-membrane interactions.  The position of the subunit-subunit interaction beads (cones) relative to the membrane excluders has little effect on the initial stages of assembly and budding, but strongly affects its completion (described in detail in section  \ref{sec:subunit} in the SI).

We note that the model does not account for local distortions within the lipid hydrophobic tails in the vicinity of the GPs. Such interactions could drive local membrane curvature and membrane-mediated subunit interactions which could either enhance or inhibit assembly and budding. Understanding these interactions is an active area (e.g. Refs. \cite{Weikl1998,Semrau2009,Reynwar2011,Goulian1993,Deserno2009}) but beyond the scope of the present study.

\subsection{Nucleocapsid}
\label{sec:NC}
The NC is represented in our model by a rigid spherical particle. This minimal representation is based on two experimental observations.  We model it as spherically symmetric because asymmetric reconstructions by Wang \etal \cite{Wang2015} showed that the alphavirus NC does not exhibit icosahedral symmetry in  virions (assembled in host cells) or viruslike particles (assembled in vitro). Second, within the NC-directed hypothesis the NC assembles completely in the endoplasmic reticulum and is then transported by the secretory pathway to the budding site at the plasma membrane. The complete NC has been shown to have a significantly higher rigidity than lipid membrane or GP-coated vesicles \cite{Schaap2012,Kol2007}; thus, we model it as infinitely rigid.

Our model NC as constructed from 623 beads distributed on a spherical surface with radius $r_{\text{NC}}=19.0\sigma$, and subjected to a rigid body constraint. To represent the hydrophobic interactions between GPs cytoplasmic tails and the capsid proteins, NC beads and the third bead of the GP subunits (counting outwards) experience an attractive Morse potential, with well-depth $\eng$. The radius of the NC sphere was tuned using bulk simulations to be commensurate with a capsid comprising 80 GPs. To minimize the number of parameters, we do not consider an attractive interaction between the NC and membrane, but the NC beads experience a repulsive Lennard-Jones potential with all membrane beads.

\subsection{Conformational change and subunit concentration}
\label{sec:conformationalmodel}
Experiments on several viral families suggest that viral proteins interconvert between `assembly-active' and `assembly-inactive' conformations, which are respectively compatible or incompatible with assembly into the virion \cite{Packianathan2010,Deshmukh2013,Zlotnick2011}. Computational modeling suggests that such conformational dynamics can suppress kinetic traps \cite{Lazaro2016,Grime2016}. Conformational changes of the alphavirus GPs E1 and E2 are required for dimerization in the cytoplasm, and it has been proposed that the GPs interconvert between assembly-inactive and assembly-active conformations \cite{Zlotnick2011}, possibly triggered by interaction with NC proteins \cite{Forsell2000}. Based on these considerations, our GP model includes interconversion between assembly-active and assembly-inactive conformations. The two conformations have identical geometries, but only assembly-active conformations experience attractive interactions to neighboring subunits. We adopt the `Induced-Fit' model of Ref.~\cite{Lazaro2016}, meaning that interaction with an assembling GP shell or the NC favors the assembly-active conformation. For simplicity, we consider the limit of infinite activation energy. In particular, with a periodicity of $\tau_\text{c}$ all the inactive subunits found within a distance 1.0$\sigma$ of the capsid are switched to the active conformation, while any active subunits further than this distance from an assembling shell convert to the inactive conformation.  Results were unchanged when we performed simulations at finite activation energies larger than $4\kt$.

In simulations performed at a constant total number of GPs  the assembly rate progressively slows over the course of the simulation due to the depletion of unassembled subunits. This is an unphysical result arising from the necessarily finite size of our simulations. Moreover, during an infection additional GPs would be targeted to and inserted into the membrane via non-equilibrium process (powered by ATP).  Therefore, our simulations are performed at constant subunit concentration within the membrane (outside of the region where an assembling shell is located). To achieve this, we include a third subunit type called `reservoir subunits', which effectively acts as a reservoir of inactive subunits. These subunits interact with membrane beads but experience no interactions with the other two types of GP subunits. With a periodicity of $\tau_\text{c}$, reservoir subunits located in a local region free of active or inactive subunits (corresponding to a circumference of radius 1.5 times the radius of the largest subunit bead) are switched to the assembly-inactive state.

\subsection{Simulations}
\label{sec:simulations}
We performed simulations in HOOMD-blue\cite{Nguyen2011}, version 1.3.1, which uses of GPUs to accelerate molecular simulations. Both the subunits and the NC were simulated using the Brownian dynamics algorithm for rigid bodies. The membrane dynamics was integrated using the NPT algorithm, a modified implementation of the Martina-Tobias-Klein thermostat-barostat. The box size changes in the membrane plane, to allow membrane relaxation and maintain a constant lateral pressure. The out-of-plane dimension was fixed at $200\sigma$.

Throughout the manuscript we report dimensions of length, mass, and energy in units of $\sigma$, $m_{0}$, and the thermal energy $\kt$.  We fixed temperature at $\kt/ \epsilon_{0}=1.1$. Physical sizes and timescales can be estimated as follows. We set the diameter of the lipid head as $\dhead= \sigma$, so that considering a 5nm-thick bilayer leads to $\sigma \approx 0.9$nm.  The characteristic timescale of the simulation is determined by the subunit diffusion, which in our simulations  is dominated by the interaction with the membrane lipids rather than with the bath. We define our unit of time $\tau_{0}$ as the characteristic time of a subunit to diffuse a distance $\sigma$ on the membrane. Comparing with a typical transmembrane protein diffusion constant $\sim 4\mu$m$^{2}$/s \cite{Ramudarai2009,Goose2013}, we obtain $\tau_{0}=250$ns.

Our simulated equations of motion do not account for hydrodynamic coupling between the membrane and the implicit solvent, which can accelerate the propagation of bilayer perturbations. To assess the significance of this effect, we performed an additional series of simulations which did account for hydrodynamic coupling, by evolving membrane dynamics according to the NPH algorithm in combination with a dissipative particle dynamics (DPD) thermostat.
As expected from Matthews and Likos \cite{Matthews2012}, we found that hydrodynamic interactions did enhance the rate of membrane deformations; however, budding proceeded only 1.1-1.2 times faster than with the NPT scheme. Moreover, the end-product distribution was the same with and without hydrodynamic interactions. Therefore, to avoid the increased computational cost associated with the DPD algorithm, we performed all subsequent simulations with the NPT method. The very limited effect of hydrodynamics can be understood from the fact that assembly timescales in our simulations are more strongly governed by subunit diffusion than by membrane dynamics (Fig. \ref{fig:timescales}).

Our system size was constrained by the capsid dimensions and the need to access long timescales. Taking the Sindbis virion as a reference structure, the bilayer neutral surface radius in the virion is $\approx 24 $nm \cite{Zhang2002}, so the surface area of the membrane envelope is $A_{0}\sim 7200$nm$^2$. We thus needed to simulate membrane patches that were significantly larger than $A_{0}$ to ensure that the membrane tension remained close to zero and that finite-size effects were negligible. Throughout this manuscript we report results from simulations on a membrane patch with size $170\times170$nm$^2$ ($A \sim 28,900$nm$^2$), which contains $51,842$ lipids. We compared membrane deformations, capsid size and organization from these simulations against a set of simulations on a larger membrane ($210\times210$nm$^2$, $A\sim44,100$nm$^2$) and observed no significant differences, suggesting that finite size effects were minimal.
Simulations were initialized with 160 subunits uniformly distributed on the membrane, including 4 active-binding subunits (located at the center of the membrane) with the remainder in the assembly-inactive conformation. In addition, there were 156 subunits in the reservoir conformation uniformly distributed. The membrane was then equilibrated to relax any unphysical effects from subunit placement by integrating the dynamics for 1,500 $\tau_{0}$ without attractive interactions between GPs.  Simulations were then performed for 4,200 $\tau_{0}$ with all interactions turned on. The timestep was set to $\Delta t=0.0015$, and the thermostat and barostat coupling constants were $\tau_{T}=0.4$ and $\tau_{P}=0.5$, respectively. Since the tension within the cell membrane during alphavirus budding is unknown, we set the reference pressure to $P_{0}=0$ to simulate a tensionless membrane. The conformational switching timescale was set to $\tau_\text{c}=3\tau_0$, sufficiently frequent that the dynamics are insensitive to changes in this parameter.
Unless otherwise specified, for each parameter set we perform 8 independent simulations.

\section{Appendix}

\subsection*{Equilibrium model for the dependence of GP shell size on system parameters}
\label{sec:EquilModel}
In this section we give a detailed derivation of Eq.~\ref{eq:nsubs} of the main text. This expression explains the simulation results for GP shell size as a function of control parameters (Fig.~\ref{fig:bending}), and is obtained  from a simple equilibrium model based on the thermodynamics of assembly \cite{Safran1994,Hagan2014} that accounts for the elasticity of the shell and the membrane.

The total free energy for the system of free subunits on the membrane, shell intermediates of size $n$, and complete shells of size $N$ can be expressed as
\begin{align}
F / \kt = \sum_{n=1}^{N} \rho_{n} [\log{\rho_{n}v_{0}}-1] + \rho_{n} \Gshell / \kt  ,
\label{eq:totalfreenergy}
\end{align}
where $\rho_{n}$ and  $\Gshell$ are respectively the concentration and interaction free energy for an intermediate with size $n$,  and $v_{0}$ is a standard state volume. Minimization of Eq.~\eqref{eq:totalfreenergy} subject to the constraint of constant subunit concentration yields the well-known law of mass action for the equilibrium distribution of intermediate concentrations,
\begin{align}
\rho_{n}v_{0}=\exp\left[-\left(\Gshell-n \mu_1\right)/ \kt \right] .
\label{eq:LMA}
\end{align}
with \begin{align}
\mu_{1} = \log{\rho_{1}v_{0}} 
\label{eq:chempotential1}
\end{align}
the chemical potential of free subunits. Similarly we can compute the chemical potential for intermediates $\mu=\partial F / \partial \rho_{n}$ as
\begin{align}
\mu_{n} = \log{\rho_{n}v_{0}} + \Gshell / \kt .
\label{eq:chempotentialn}
\end{align}
For large shells, the first term in \eqref{eq:chempotentialn} is neglegible compared to the  free energy of the shell, and the chemical pontential can be approximated as $\mu_{n}\approx \Gshell/ \kt$. In equilibrium, the chemical potential of free subunits must be equal to that of subunits in shells and intermediates, leading to $\mu_{1}=\mu_{n} / n \approx \Gshell / n\kt$.

The intermediate size with maximal concentration is determined by the condition
\begin{align}
\frac{d\rho_{n}}{dn}=\rho_{n}\frac{d}{dn}[-\Gshell \kt + n \mu_{1}] =0,
\label{eq:rhomax}
\end{align}
which using \eqref{eq:chempotential1}  and \eqref{eq:chempotentialn} can be rewriten as
\begin{align}
\frac{d}{dn}[-\Gshell/ \kt + n \mu_{1}] \approx \left[- \frac{d \Gshell}{dn} + \Gshell / n \right]/ \kt.
\label{eq:rhomax2}
\end{align}
Thus, the optimal size at equilibrium is that which minimizes the interaction free energy per subunit, $\Gshell/n$.

The interaction free energy includes subunit-subunit interactions and the elastic energy of the shell and the membrane. Assuming that the shell can be described as a continuous, two-dimensional spherical shell, its elastic energy is given by the Helfrich bending energy, with bending modulus $\kcap$ and spontaneous curvature $c_{0}=2/R_{0}$, where $R_{0}$ is the equilibrium radius of the shell. The membrane underneath is a symmetric bilayer with rigidity $\kmem$. The free energy $\Gshell$ thus reads
\begin{align}
\Gshell = n \Delta g_{\text{g}} + \frac{\kmem}{2} \int_{S} c^{2} dS +  \frac{\kcap}{2} \int_{S} (c-c_0)^{2} dS,
\label{eq:helfrich}
\end{align}
where $\Delta g_{\text{g}}$ is the free energy per subunit added to the shell,  $c$ is the total curvature, and $S$ denotes the surface area. Assuming spherical symmetry and accounting for the fact that the subunits are rigid, the shell surface area is $S \approx n S_{0}$, with $S_{0}$ as the area per subunit. We can then express the total curvature as a function of the number of subunits in the particle, $c= 2/R \approx (16\pi / n A_{0})^{1/2}$.
Here we are assuming that the Gaussian modulus of the membrane is unchanged by the presence of the GPs, so that the integrated Gaussian curvature is constant for fixed topology by the Gauss-Bonnet theorem \cite{Deserno2015}. Under the (reasonable) assumption that the shell size is determined before scission, the Gaussian curvature energy then contributes a constant to the free energy and can be neglected. We have also neglected the energy from the 12 disclinations in the shell. Accounting for this could shift the theory curve in Fig.~\ref{fig:bending}b  but would not change the slope.

Finally, recalling that the equilibrium configuration minimizes the interaction free energy per subunit, $\Gshell/n$ results in \eqref{eq:nsubs} of the main text,
\begin{align}
n = n_{0} \left( 1 + \frac{\kmem}{ \kcap} \right)^{2},
\label{eq:nsubs}
\end{align}
where $n_{0}$ is the number of subunits in the equilibrium configuration in the absence of membrane, corresponding to $n_{0}=80$ in our model.

\subsection*{Acknowledgements}
This work was supported by the NIH, Award Number R01GM108021 from the National Institute Of General Medical Sciences (GRL and MFH), the Brandeis Center for Bioinspired Soft Materials, an NSF MRSEC,  DMR-1420382 (GRL), and the NSF, award number MCB1157716 (SM). Computational resources were provided by the NSF through XSEDE computing resources (MCB090163) and the Brandeis HPCC which is partially supported by the Brandeis MRSEC.

\bibliography{all-references-alphavirus}

\newpage

\section*{SUPPLEMENTAL INFORMATION}
\setcounter{figure}{0}
\setcounter{section}{0}
\renewcommand{\thefigure}{S\arabic{figure}}
\renewcommand{\thesection}{S\arabic{section}}
\setcounter{equation}{0}
\renewcommand{\theequation}{S\arabic{equation}}

\section{Assembly and budding without conformational switching}
\label{sec:conformational}
Fig. \ref{fig:conformational} compares the fraction of trajectories leading to complete GPNC particles with and without conformational switching as a function of the GP-GP interaction strength. We see that, in the absence of conformational switching, complete assembly and budding only occurs in the limit of weak GP-GP interactions and a strong GP-NC interaction. Stronger GP-GP interactions allow nucleation of GP shells away from the vicinity of the nucleocapsid, thus leading to a kinetic trap in which too many shells have nucleated (Fig. \ref{fig:conformational}b). Moderate GP-GP interactions avoid this trap, leading to assembly and budding of well-formed shells in about half of simulated trajectories. However, when the GP-GP interaction is further decreased, the assembly trajectory is dominated by the NC interaction. The GP subunits adsorb onto the NC without forming a well-defined lattice, leading to defective particles with holes. An example of such a configuration is shown within the plot.

 \begin{figure}[hbt]
  \begin{center}
  \includegraphics[width=\columnwidth]{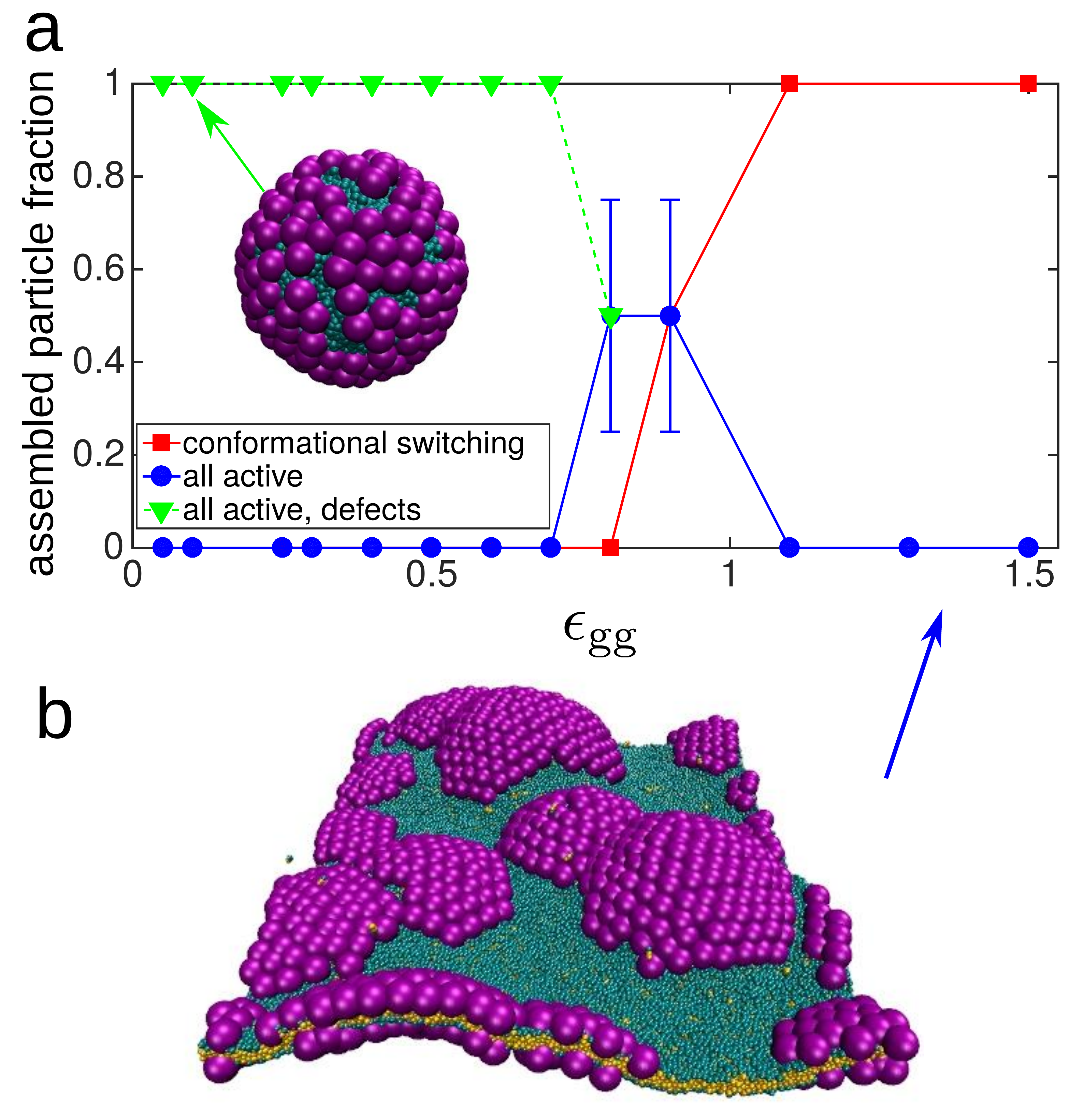}
    \caption{ Effect of conformational switching (CS) on assembly and budding around a nucleocapsid. \textbf{a}  Fraction of trajectories in which a complete shell assembled as a function of the GP-GP interaction strength, with constant $\eng = 3.5$. Results are shown for simulations in which all the subunits are active (`all active', \textcolor{blue}{$\CIRCLE$}) and simulations with conformational switching (`conformationa switching', \textcolor{red}{$\blacksquare$}). The fraction of trajectories leading to shells with large holes is shown for the case with conformational switching (`all active, defects', \textcolor{green}{$\blacktriangledown$}). Each data point corresponds to 4 independent simulations. \textbf{b} Snapshot showing a typical, kinetically trapped configuration from a simulation without conformational switching with $\egg=1.3$.}
  \label{fig:conformational}
  \end{center}
\end{figure}

\section{Subunit geometry and membrane reshaping}
\label{sec:subunit}

\begin{figure}[hbt]
 \begin{center}
  \includegraphics[width=0.8\columnwidth]{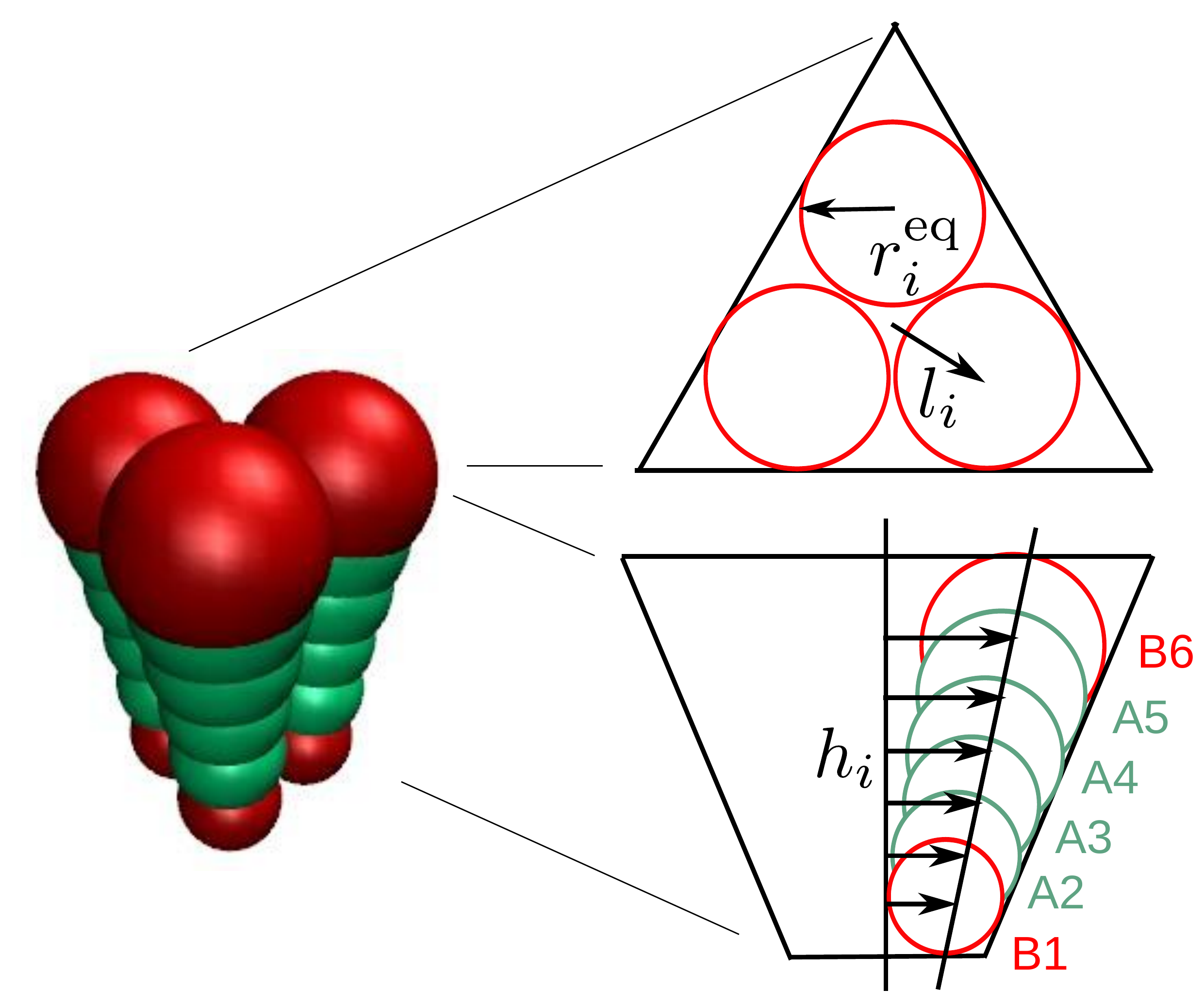}
    \caption{ Schematic of the subunit geometry, with views from directly above the plane of the membrane and within the plane of the membrane. Membrane excluders are not shown in these schematics to aid visual clarity. }
  \label{fig:subunitgeom}
  \end{center}
\end{figure}

\emph{Subunit geometry.} The geometry of the model GP trimer subunit used in the main text is schematically shown in Fig. \ref{fig:subunitgeom}. As explained in the main text, the subunit consists of three cones symmetrically placed around the subunit axis. Each cone contains six pseudoatoms. Only the inner four pseudoatoms (denoted as A) experience attrative interactions. The outer two pseudoatoms, B, interact with the rest through excluded volume. The pseudoatoms are placed at heights $h_{i}=[16.0,17.5,19.0,20.5,22.0,23.5]\sigma$. At each plane $z=h_{i}$ there are three identical pseudoatoms forming an equilateral triangle of radius $\li=h_{i}\tan{\al}$, where $\al$ can be tuned. Since assembly in bulk is slightly more robust for smaller $\al$, we choose an optimal value $\al=7^{\circ}$.  The radius of each pseudoatom is then given by $\reqi=\li\cos{\psi}$, being $\psi$ the parameter that controls the preferred curvature of the subunits. We set $\psi=94.9^{\circ}$ (see section \ref{sec:bulk}). Finally, to embed the subunits in the membrane we add two layers of three membrane excluders ` VX ', consistent with the cone geometry, at height $h_{\text{in}}=19.0\sigma$ (inner domain) and $h_{\text{out}}=26.0\sigma$ (outer domain). The sequence of pseudoatoms across the shell reads [B$_{1}$,A$_{2}$,A$_{3}$,VX$_{\text{in}}$,A$_{4}$,A$_{5}$,B$_{6}$,VX$_{\text{out}}$].

\emph{Effect of changing the subunit geometry.} In our initial model for GP subunits, the cones were positioned entirely within the membrane, such that all lateral interactions between neighboring cones were within the body of the membrane (Fig. \ref{fig:neckgeometry}). Specifically, the membrane excluders were located at the same positions as the pseudoatoms B$_{1}$ and B$_{6}$.  However, this subunit structure led to the formation of short budding necks around partial GP shells. The high curvature within such necks presented an extremely large barrier to subunit diffusion, and hence prohibited complete assembly and budding.

We explored several other subunit geometries. We found that displacing the cone attractor beads further to the outside of the membrane leads to similar results. However, when the cone pseudoatoms are displaced towards the interior of the shell, so that the lateral interactions take place below the membrane, a longer neck with more shallow curvature forms. The reduced neck curvature lowers the barrier to subunit diffusion, allowing complete assembly and budding. The difference in neck geometry likely arises because the lower position of the subunit attractions allows them to exert a higher torque on the membrane. We note that this particular aspect of our subunit geometry does not conform to the actual GP structure and interactions; the lateral attractions between Sindbis GPs are primarily situated above the membrane. However, the neck geometry in these simulations (long and with shallow curvature) closely resembles those observed in experiments of in vivo virus assembly.

\begin{figure}[hbt]
  \begin{center}
  \includegraphics[width=\columnwidth]{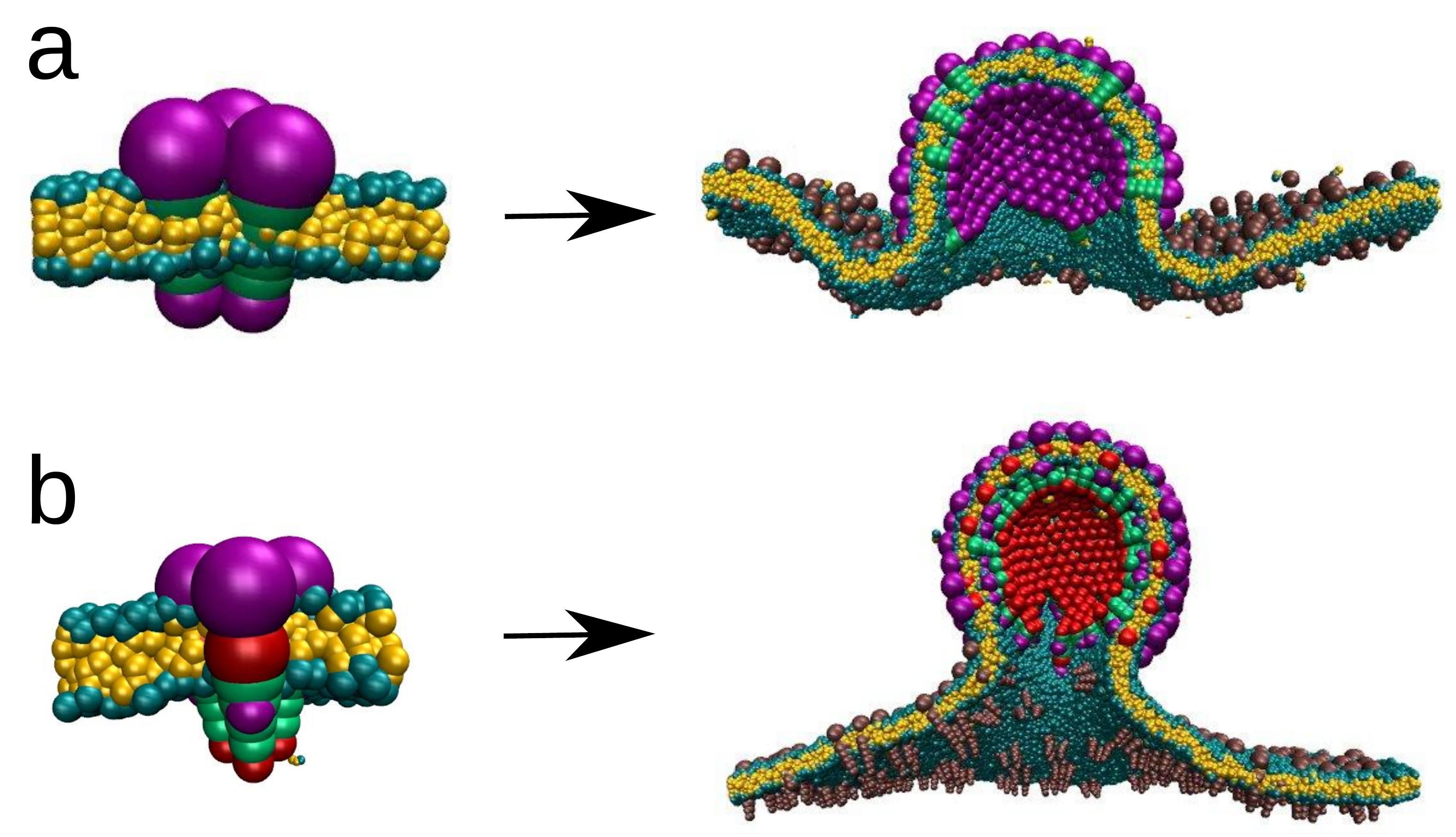}
    \caption{ Relationship between the structure of glycoproteins and their ability to reshape the membrane. \textit{(Top)} When `A' pseudoatoms overlap with the membrane and the membrane excluders overlap with `B' pseudoatoms, the budding neck develops acute angles around a partially assembled shell. This leads to a large barrier to subunit diffusion that prevents complete assembly and budding.   \textit{(Bottom)} When both `A' and `B' pseudoatoms are situated below the membrane, the budding neck is longer, with a shallower angle. This reduces the barrier to subunit diffusion, allowing completion of assembly and budding.  }
  \label{fig:neckgeometry}
  \end{center}
\end{figure}

\begin{figure}[hbt]
  \begin{center}
  \includegraphics[width=\columnwidth]{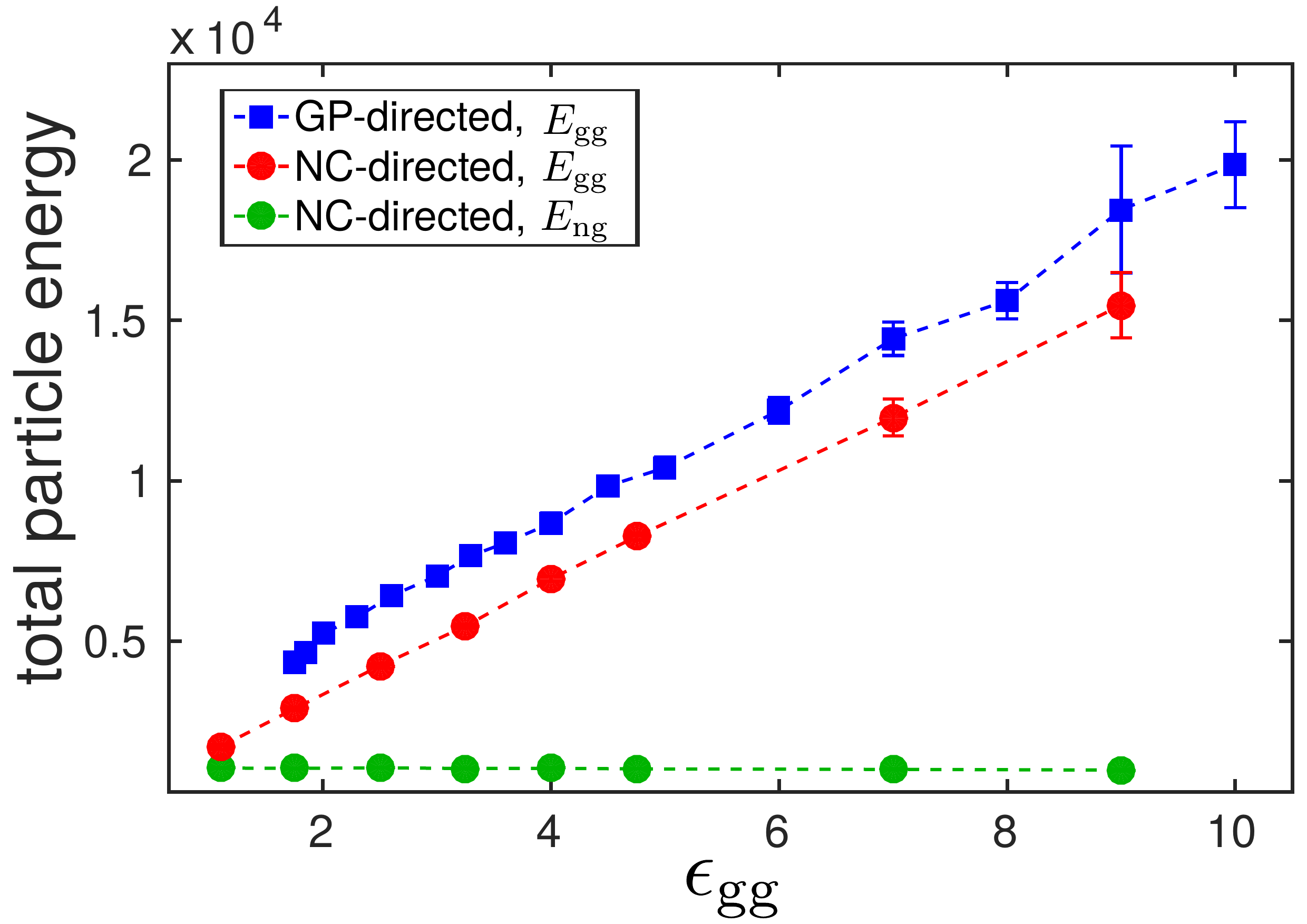}
    \caption{ Total particle energy, accounting for the attractive energy of the Morse potentials of all the pseudoatoms that form the particle, as a function of the subunit interaction $\egg$. Results are shown for a GP-particle (\textcolor{blue}{$\blacksquare$} symbols) and a GPNC-particle with $\eng=3.5$. For the GPNC-particle, the energy is separated into the components arising from GP-GP interactions ($E_{\text{gg}}$, \textcolor{dlgreen}{$\CIRCLE$} symbols) and from GP-NC interactions ($E_{\text{ng}}$, \textcolor{red}{$\CIRCLE$} symbols).}
  \label{fig:energy}
  \end{center}
\end{figure}

\begin{figure}[hbt]
  \begin{center}
  \includegraphics[width=\columnwidth]{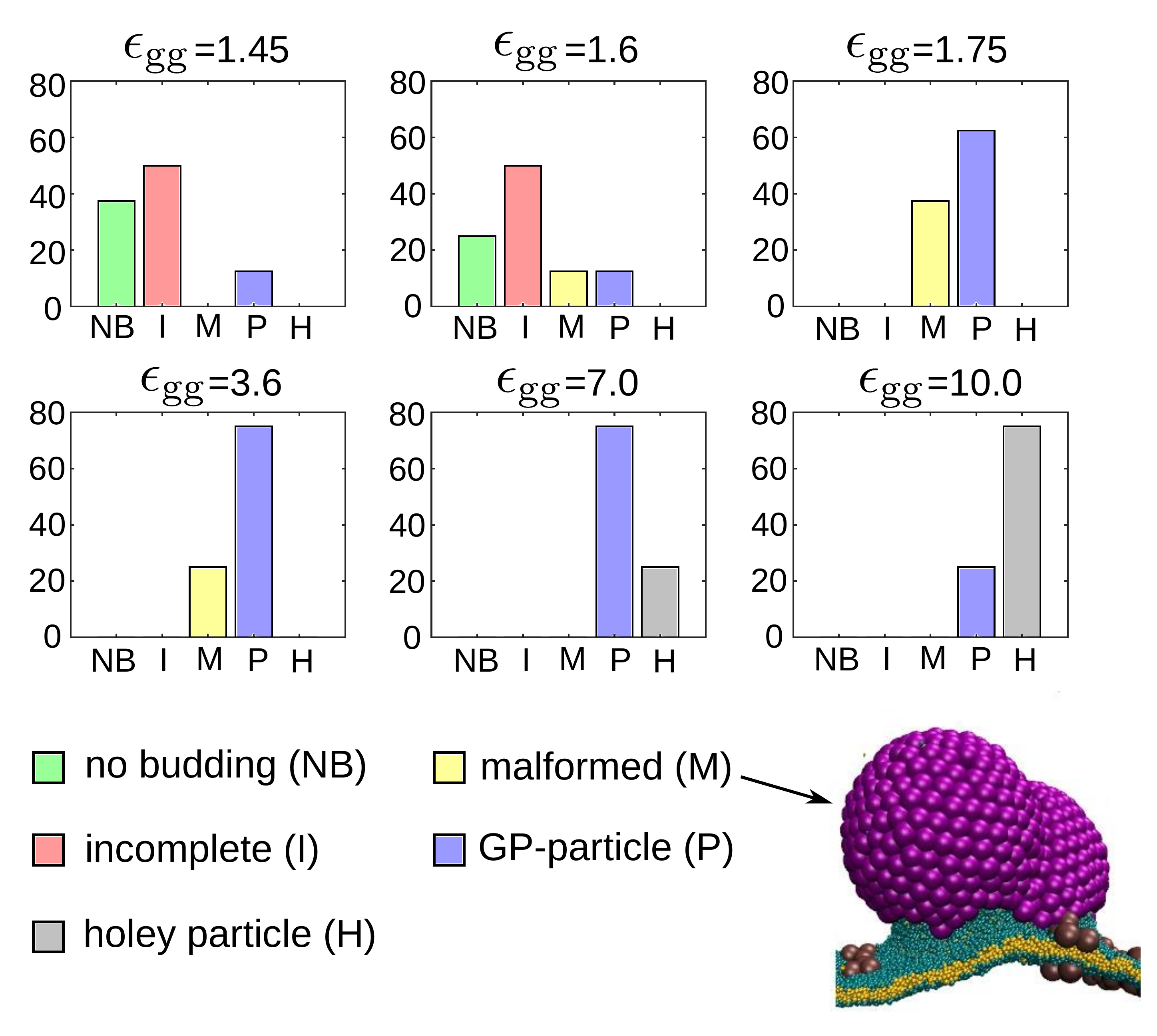}
    \caption{Probability of each assembly outcome for some representative values of $\egg$ in the GP-directed budding. We find no parameter value for which malformed capsids (for which a typical example is shown) are predominant, but they represent up to 40\% of the simulation outcomes at $\egg=1.75$. }
  \label{fig:distributionoutcomes}
  \end{center}
\end{figure}

\section{Interaction potentials}
\label{sec:potentials}

The total interaction energy $U_{\text{tot}}$ can be separated into three contributions,

\begin{align}
U_{\text{tot}} =  U_{\text{mem}} + U_{\text{gg}} +  U_{\text{nc}},
\label{eq:Utot}
\end{align}

\noindent where $U_{\text{mem}}$ represents the interaction energy between the membrane beads, $U_{\text{gg}}$ accounts for the interaction of between subunits as well as with the membrane, and $U_{\text{nc}}$ represents the interaction energy of the NC with the subunits and the membrane.

\subsection{Membrane interactions}
The membrane lipids consist of three beads, the first representing the lipid head and the other two  connected through two finite extensible nonlinear elastic (FENE) bonds with maximum length $r_{\text{cut}}=1.5\sigma$,

\begin{align}
U_{\text{bond}}(r)=-\frac{1}{2} k_{\text{bond}}r_{\text{cut}}^{2} \log{[1-(r/r_{\text{cut}})^{2}]}.
\end{align}

\noindent with $k_{\text{bond}}=30\epsilon_{0}/\sigma^{2}$. A harmonic spring links the two outer beads, to ensure that the lipids maintain a cylindrical shape,

\begin{align}
U_{\text{bend}}(r)=\frac{1}{2} k_{\text{bend}}(r-4\sigma)^{2}.
\end{align}

All membrane beads interact via a Weeks-Chandler-Andersen potential,

\begin{align}
U_{\text{rep}}(r)=\sum 4\epsilon_{\text{rep}} \left[ \left( \frac{b_{i,j}}{r} \right)^{12} - \left( \frac{b_{i,j}}{r} \right)^{6}  + \frac{1}{4} \right],
\end{align}

\noindent with $\epsilon_{\text{rep}}=1$ and cutoff $r_{\text{cut}}=2^{1/6}b_{i,j}$. The parameter $b_{i,j}$ depends on the identities of the interacting beads: $b_{\text{h,h}}=b_{\text{h,t}}=0.95\sigma$ and $b_{\text{t,t}}=1.0\sigma$, with the subscripts `h' and `t' denoting head and tail beads, respectively.
The hydrophobic nature of the lipid tails is accounted for by an attractive interaction between all pairs of tail beads:

\begin{equation}
U_{\text{hydro}}(r)=
\begin{cases}
-\epsilon_{0},  & r < r_\text{c}  \\
-\epsilon_{0}\cos{[\pi(r-r_\text{c})/2\omega_\text{c}]}, &  r_\text{c} \leq r \leq r_\text{c} + \omega_\text{c} \\
0,    & r > r_\text{c}+\omega_\text{c}
\end{cases}
\end{equation}

\noindent with $\epsilon_{0}=1.0$, $r_\text{c}=2^{1/6}\sigma$. The potential width $\omega_\text{c}$ is a control parameter that determines, among other properties, the membrane rigidity. Unless otherwise specified, $\omega_\text{c}=1.6$.

\subsection{GP-GP interactions}
The interaction potential between GP subunits, $U_{\text{gg}}$, consists of two terms. The attractive interaction between a pair of attractor pseudoatoms `A' of the active subunits is modeled by a Morse potential. Beads interact only with those of the same kind on a neighboring cone, A$_{i}$-A$_{i}$, $i=2,..,5$, and the equilibrium distance of the potential depends on the pseudoatom radius, $r^\text{eq}_{i}$:
\begin{align}
U_{\text{gg}}^{M}= \sum_{i=2}^{5} U_{i}^{M}= \sum_{i=2}^{5} \egg ( e^{-2\alpha_{i}(r-2\reqi)}) -2e^{-\alpha_{i}(r-2\reqi)})
\label{eq:Ugg}
\end{align}

\noindent with $\alpha_{i}=(3.0/\reqi)$. The cutoff of this interaction was set at $r_{\text{cut}}=2r^{eq}_{i}+3.5$.  The subunit beads also experience excluded volume interactions,
\begin{align}
U_{\text{g-g}}^{\text{ex}}(r)= \sum_{i} \sum_{j} 4\epsilon_{\text{ex}} \left[ \left( \frac{b_{i,j}}{r} \right)^{12} - \left( \frac{b_{i,j}}{r} \right)^{6} \right]
\label{eq:LJrep}
\end{align}
 with $\epsilon_{\text{ex}}=1.0$ and cutoff radius $r_{\text{cut}}=b_{ij}=\reqi+r^{\text{eq}}_{j}$. The sum extends to all the subunit beads, both active and inactive.

In the subunits, only the pseudoatoms `VX' interact with the membrane beads; there is no interaction between membrane beads and `A' or `B' pseudoatoms. The interaction between subunit excluders and membrane beads corresponds to the repulsive part of the Lennard-Jones potential,

\begin{align}
U_{\text{g-m}}^{\text{ex}}(r)= \sum_{i} \sum_{j} 4\epsilon_{\text{ex}} \left[ \left( \frac{b_{i,j}^{\text{g-m}}}{r} \right)^{12} - \left( \frac{b_{i,j}^{\text{g-m}}}{r} \right)^{6} \right],
\label{eq:LJrep2}
\end{align}

\noindent where $i$ runs over all lipid beads and $j$ over all `VX' pseudoatoms, and $b_{i,j}^{\text{g-m}}=0.5+r_{\text{in}}$ for the inner excluders VX$_{\text{in}}$ and $b_{i,j}=0.5+r_{\text{in}}$ for the outer excluders VX$_{\text{out}}$.

\subsection{NC interactions}

The NC beads interact with pseudoatoms A$_{3}$ of the active GP subunits through a Morse potential,
\begin{align}
\UggM(r)= \sum_{\text{nc, j}} \sum_{\text{A}_{3}}  \eng ( e^{-2\alpha_{\text{nc}}(r-2r^{\text{eq}}_{\text{nc}})}) -2e^{-\alpha_{\text{nc}}(r-2r^{\text{eq}}_{\text{nc}})})
\label{eq:Ugg}
\end{align}
 with $r^{\text{eq}}_{\text{nc}}=1.0$ and $\alpha_{\text{nc}}=2.5$. We explored a broad range of $\alpha_{nc}=1.5-6.0$ and observed little difference in the morphology of the assembly product.

NC beads interact with all the membrane beads through a repulsive Lennard-Jones potential,

\begin{align}
U_{\text{nc-m}}^{\text{ex}}(r)= \sum_{i} \sum_{j} 4\epsilon_{\text{ex}} \left[ \left( \frac{b_{i,j}^{\text{nc-m}}}{r} \right)^{12} - \left( \frac{b_{i,j}^{\text{nc-m}}}{r} \right)^{6} \right],
\label{eq:LJrep3}
\end{align}

\noindent where $i$ runs over all lipid beads and $j$ over all NC pseudoatoms, and $r_{\text{cut}}=b_{i,j}^{\text{nc-m}}=1.0$. Similarly, there is a a repulsive Lennard-Jones potential between NC pseudoatoms and certain GP pseudoatoms

\begin{align}
U_{\text{nc-g}}^{\text{ex}}(r)= \sum_{i} \sum_{j} 4\epsilon_{\text{ex}} \left[ \left( \frac{b_{i,j}^{\text{nc-g}}}{r} \right)^{12} - \left( \frac{b_{i,j}^{\text{nc-g}}}{r} \right)^{6} \right],
\label{eq:LJrep4}
\end{align}

\noindent where $i$ runs over all NC beads and $j$ over all the subunit pseudoatoms of the type  A$_{4}$, A$_{5}$, B$_{6}$ and VX$_{\text{out}}$. We set the distance $b_{ij}^{\text{nc-g}}=0.5+r_{i}^{\text{eq}}$, with 0.5 the radius of the NC beads.

\section{Bulk simulations}
\label{sec:bulk}
The preferred curvature of the GP shell can be tuned by varying the cone angle $\psi$, which leads to aggregates of different shapes and sizes. To determine the relationship between cone angle and aggregate size, we performed bulk simulations (i.e. without a membrane present) of GP assembly. In these simulations we initialized 200 subunits with random positions and orientations (except not overlapping) in a box of size 180$^{3}\sigma^{3}$. The subunit interaction was set at $\egg=0.995$, which allows assembly with high yield as shown in Fig. \ref{fig:capsidsbulk}a. We find that for the cone angle $\psi=94.9^{\circ}$ the predominant assembled shell is roughly spherical and contains 80 subunits. The distribution of assembly products is shown in Fig. \ref{fig:capsidsbulk}b for 19 simulations. The subunits exhibit hexagonal packing on the shell surface, and 5-fold disclinations can be identified. However, in general, the spatial distribution of the defects is not consistent with icosahedral organization.

We also performed bulk simulations examining GP assembly around a NC. These simulations were the same as described in the previous paragraph, except that the simulation box included one NC particle. We performed simulations over a range of NC radii to determine the optimal size for assembly of the GP shell, which identified an optimal NC radius of $r_{\text{NC}}=19.0\sigma$. This value is consistent with the position of the contact between the GP and the NC in Sindbis virus, $\sim 19.5$nm \cite{Zhang2002}. The distribution of end-products for bulk simulations at this NC radius are shown in Fig. \ref{fig:capsidsbulk}b.

 \begin{figure}[hbt]
  \begin{center}
  \includegraphics[width=\columnwidth]{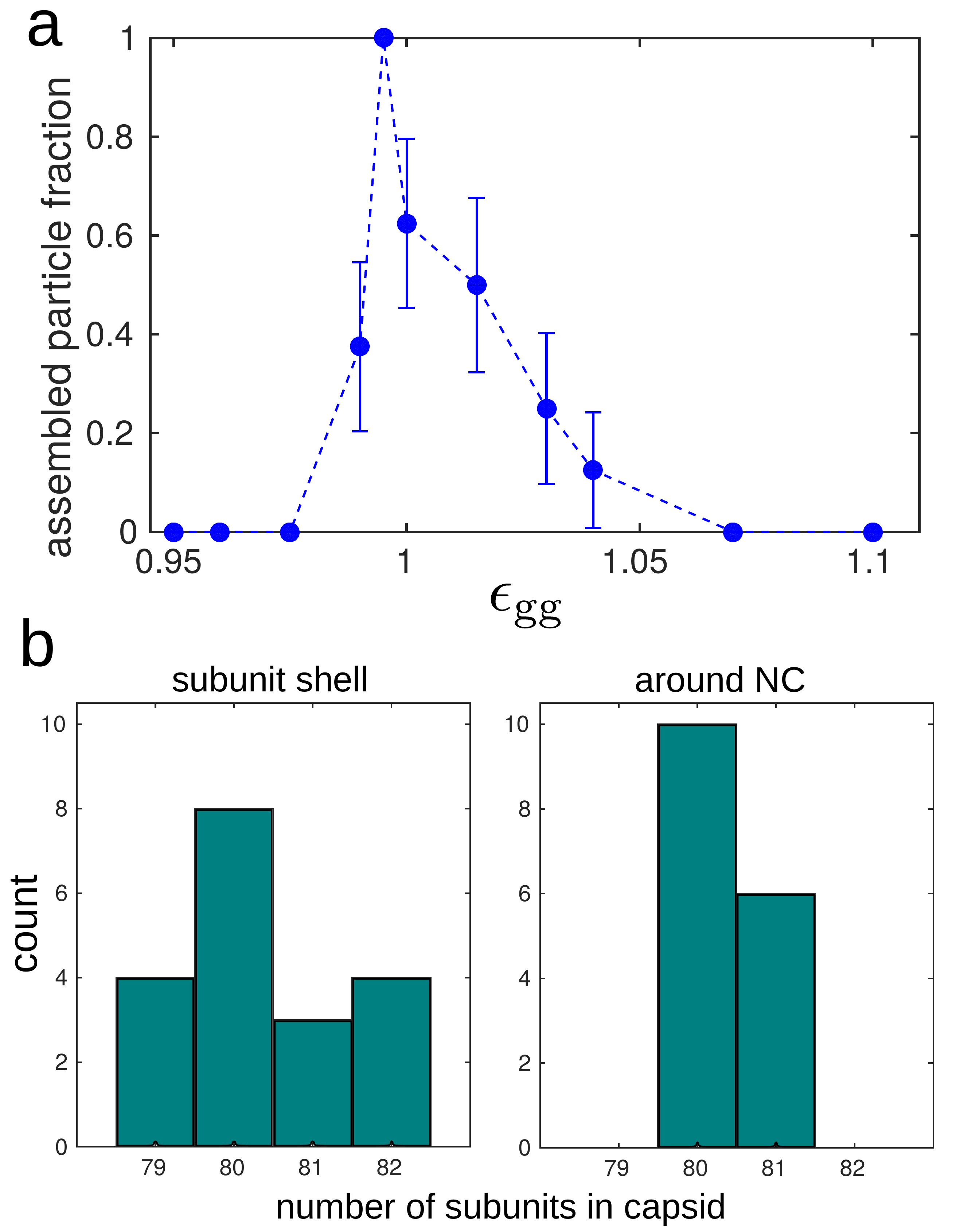}
    \caption{ \textbf{a} Fraction of trajectories that lead to assembled particles as a function of subunit interaction $\egg$, for 8 independent simulations. Parameters are: 200 GP subunits in a box with side length 180$\sigma$, with each simulation performed for $8.75\cdot 10^{5}\tau_0$. \textbf{b} Distribution of assembly products in bulk simulations (\ie in the absence of the membrane). \textit{(Left)} Assembly of GP subunits without a NC. \textit{(Right)} Assembly around a spherical NC. In both cases $\egg=0.995$, and $\eng=1.8$ for assembly around a NC. Each distribution is calculated from 19 independent simulations.  }
  \label{fig:capsidsbulk}
  \end{center}
\end{figure}

\section{Shell and membrane bending modulus estimation}
\label{sec:bendingestimation}
\subsection{Shell bending modulus}

Our estimation of the shell bending modulus  is based on the work on triangulated surfaces by \citet{Gompper1996}. The discretization of the curvature in terms of the squared difference of the normal vector of neighboring subunits allows to express the discrete Helfrich bending energy as

\begin{align}
 H_{\text{B}} = \frac{k}{2} \sum_{\alpha,\beta} (\hat{n}_{\alpha} - \hat{n}_{\beta})^{2} \equiv k \sum_{\alpha,\beta} (1 -\hat{n}_{\alpha} \cdot \hat{n}_{\beta}),
\label{eq:kappa1}
\end{align}

\noindent with $k=\sqrt{3}\kappa$ and $\kappa$ as the bending modulus. $\hat{n}_{\alpha}$ represents the normal vector to the subunit $\alpha$, so that the angle $\theta$ between two adjacent subunits is given by $\hat{n}_{\alpha} \cdot \hat{n}_{\beta}=\cos{\theta}$. The energy can be rewriten as

\begin{align}
 H_{\text{B}}  = k \sum_{\alpha,\beta} [1 - \cos(\theta - \theta_{0})],
\label{eq:kappa2}
\end{align}

\noindent where $\theta_{0}$ corresponds to the preferred curvature of the lattice, and the sum runs over all the subunit pairs interacting in the shell. Assuming small variations around the preferred angle results in
\begin{align}
 H_{\text{B}} \approx \frac{k}{2} \sum_{\alpha,\beta} (\theta - \theta_{0})^{2}.
\label{eq:kappa3}
\end{align}

 Therefore, if the interaction energy $U_{\text{gg}}$ between subunits can be expressed as a function of the angle $\theta$, comparison with \eqref{eq:kappa3} allows to estimate the bending rigidty in terms of the parameters that define the interaction. As opposed to the two-dimensional case of triangulated surfaces, our subunits are three dimensional structures of finite thickness. The interaction between subunits is given by the Morse potential between `A' pseudoatoms,

 \begin{align}
 U_{\text{gg}}^{M} = \sum_{i} \UMi(r_{i}),
\label{eq:kappa4}
\end{align}

\noindent where $r_{i}$ is the distance between pseudoatoms of the same type, and the index $i$  runs over the four pseudoatoms $\{A_{i}\}$. In equilibrium, the angle between subunits is given by $\theta_{0}$. But if the shell is subject to mechanical perturbations, the subunits will tilt around the neutral surface with an angle $\theta$. For small perturbations, the potential \eqref{eq:kappa4} is approximated by
\begin{align}
 \UggM \approx \frac{1}{2}\frac{\partial^{2}\UggM}{\partial \theta^{2}} \Bigr|_{\theta=\theta_{0}}(\theta - \theta_{0})^{2} .
\label{eq:kappa5}
\end{align}

 \begin{figure}[hbt]
  \begin{center}
  \includegraphics[width=\columnwidth]{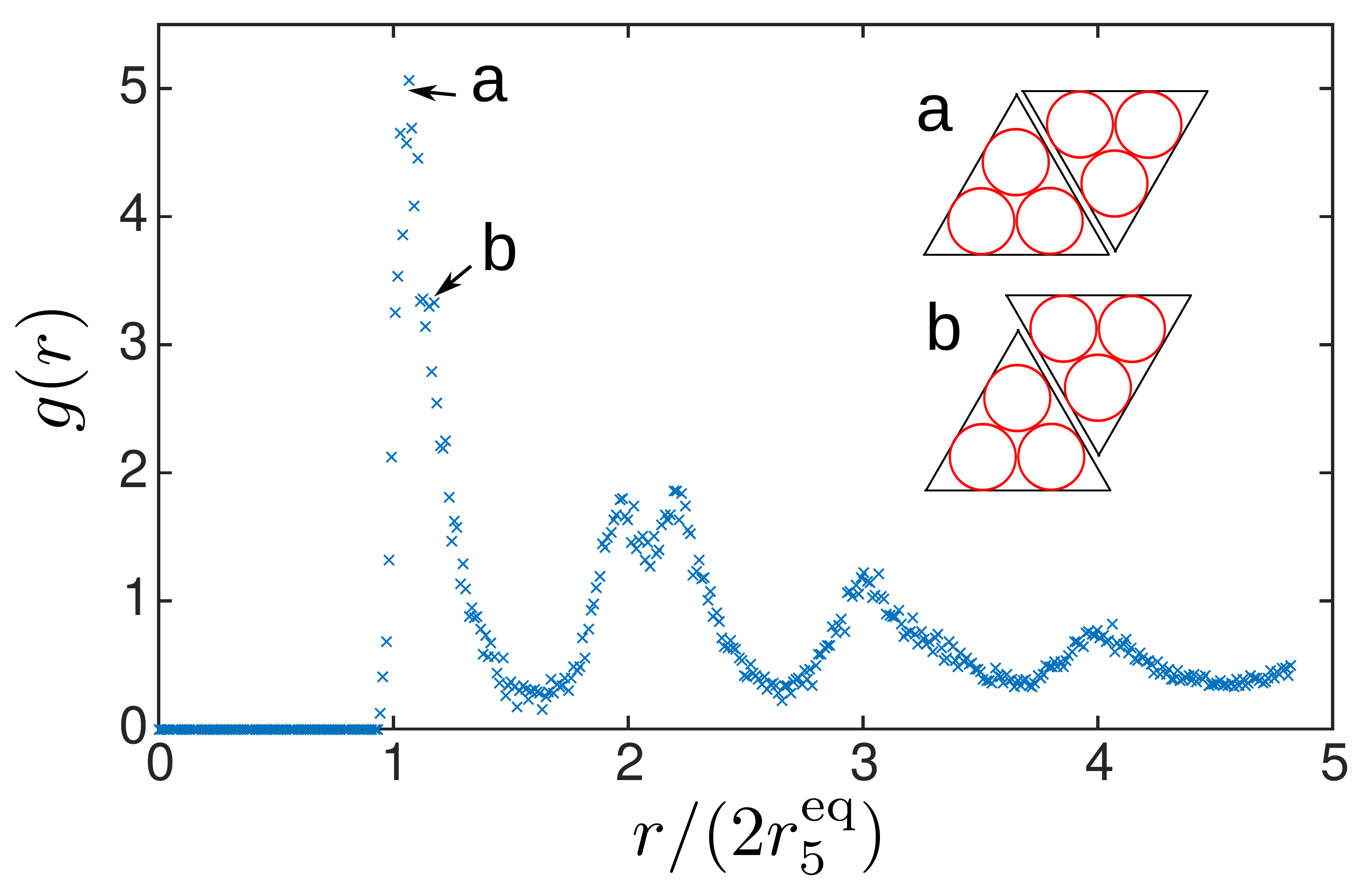}
    \caption{ Pair correlation function $g(r)$ for cone positions within assembled GP shells. To compute $g(r)$ we measure separations between pairs of A$_{5}$ pseudoatoms. In the plot we have scaled the separation $r$ by the equilibrium distance of the Morse potential. The correlation function suggests that two possible configurations are present in the shell. In the configuration \textbf{a}, the nearest neighbor is located at a distance $r_{\textbf{a}}=2r^{\text{eq}}_{5}$, whereas in configuration  \textbf{b} it is found at $r_{\textbf{b}}=\sqrt{3l^{2}_{5}/4+(2r_{5}^{\text{eq}})^{2}}\approx 1.16r_{\textbf{a}}$.}
  \label{fig:paircorrelation}
  \end{center}
\end{figure}

To determine the equilibrium positions of the subunits in the shell, we need to assess how they organize in the shell. Although the resulting structures are not perfectly ordered and there is some variation in the distribution of subunits,  analysis of the pair correlation function $g(r)$, shown in Fig. \ref{fig:paircorrelation}, suggests that subunits organize into two configurations. We assume that the configuration in which the cones are separated by the equilibrium distance of the Morse potential, Fig. \ref{fig:paircorrelation}a, is dominant. In this configuration, the angle that minimizes the interaction energy $U_{\text{gg}}$ is the angle that minimizes separately each term of \eqref{eq:kappa4}. By comparing expressions \eqref{eq:kappa3} and \eqref{eq:kappa5}, the shell bending modulus can be expressed as

\begin{align}
\kcap = \frac{1}{\sqrt{3}} \sum_{\alpha, \beta} \sum_{i} \frac{\partial^{2}\UMi}{\partial^{2} \theta} \Bigr|_{\theta=\theta_{0}},
\label{eq:kappa6}
\end{align}

A more detailed explanation of the interaction is given in Fig. \ref{fig:sketchelasticity}. In this configuration, the pseudoatoms in cone `a' in the subunit $\alpha$ interact with pseudoatoms in cones `b' and `c' in subunit $\beta$ and the analogous pseudoatoms in subunit $\gamma$. Taking into acocunt the symmetry of the system, we only consider the interaction of the cone `a' with cones `b' and `c'. Pseudoatoms A$_{i}$ in cones 'a' and 'b' are separated by a distance $r^{\text{ab}}_{i}=2r_{i}^{\text{eq}}$. Pseudoatoms `a' and `c' are however separated by a much larger distance, $r^{ac}_{i}=\sqrt{(r^{\text{ab}}_{i})^{2}+3\li^{2}}$. Hence we only consider interactions between nearest neighbors and neglect second-neighbors.

The Morse potential explicitly depends on the distance between pseudoatoms, so in practice one needs to express this distance in terms of the angle, $r_{i}=r_{i}(\theta)$, and thus the previous expression reads

\begin{align}
\kcap = \frac{1}{\sqrt{3}}  \sum_{\alpha, \beta} \sum_{i} \frac{\partial^{2}\UMi}{\partial r_{i}^{2}} \left( \frac{\partial r_{i}}{\partial \theta}\right)^{2} \Bigr|_{\theta=\theta_{0}},
\label{eq:kappa7}
\end{align}

\noindent where the second derivative of the Morse potential yields

\begin{align}
 \frac{\partial^{2}\UMi}{\partial r^{2}} \Bigr|_{r=2r_{i}^{\text{eq}}} = 2\egg \alpha^2_{i},
\label{eq:der2U}
\end{align}

Considering the symmetry of the system, we can compute the total interaction as $3n_{\text{i}}/2$ the interaction between a pair of cones, where 3 corresponds to the number of cones per subunit and $n_{\text{i}}$ is the average number of interacting neighbor cones. Consistent with hexagonal packing, we set $n_{\text{i}}=4$ (since two neighbor cones are in the same rigid body).

The neutral surface of the interaction between subunits corresponds to the position at which the stress between the subunits vanish, given by the condition $d\UggM/dx=0$ along the midsurface between both subunits. We calculated the position of the neutral surface $h_{n}$ numerically, and found a position close to central point of the cones, situated between pseudoatoms A$_{3}$ and A$_{4}$. For simplicity, we take $h_{\text{n}}=(h_{3}+h_{4})/2$.

To obtain the distance between the pseudoatoms in cones `a' and `b' with respect to variations in the angle between subunits when they rotate with respect to the neutral surface, we initialize the subunits centered at the neutral surface position. Following the scheme shown in Fig. \ref{fig:sketchelasticity} the coordinates of the pseudatoms in cone `a' and `b' are initially given by

\begin{eqnarray}
\vec{a}_{i}=[\li/2,\sqrt{3} \li/2,\bar{h}_{i}], \\
\vec{b}_{i}=[-\li/2,\sqrt{3} \li/2,\bar{h}_{i}],
\label{eq:kappa9}
\end{eqnarray}

\noindent where we have introduced $\bar{h}_{i}=h_{i}-h_{\text{n}}$. The subunits are rotated by an angle $\theta/2$ in the case of `a' and by $-\theta/2$ in `b'. Subunits $\alpha$ and $\beta$ are then translated by a distance $-d_{\text{n}}/2$ and $+d_{\text{n}}/2$, respectively, along the direction $\hat{x}$, with $d_{\text{n}}$ as the equilibrium distance between subunits at a height $h_{\text{n}}$.

\begin{eqnarray}
\vec{a}_{i} \rightarrow \hat{\mathcal{R}}_{y}(\theta/2)\vec{a}_{i}-d_{n}/2 \hat{x}. \\
\vec{b}_{i} \rightarrow \hat{\mathcal{R}}_{y}(-\theta/2)\vec{a}_{i}+d_{n}/2 \hat{x}.
\label{eq:kappa10}
\end{eqnarray}

Finally, the distance between pseudoatoms a$_{i}$ and b$_{i}$  reads

\begin{align}
r_{i}^{\text{ab}}(\theta)=2\bar{h}_{i}\sin(\theta/2)-\li\cos(\theta/2)+d_{\text{n}}.
\label{eq:kappa11}
\end{align}

As suggested by analysis of Fig. \ref{fig:paircorrelation}, in the shell the distance between atoms is given by the equilibrium distance of the Morse potential, $2\reqi$. Taking $r_{i}^{\text{ab}}=2\reqi$, this set of equations allows to evaluate the equilibrium angle $\theta_{0}$ and distance between the subunits $d_{\text{n}}$, obtaining $\theta_{0}=18.6^{\circ}$ and $d_{\text{n}}=7.36\sigma$. Computing the derivative of expression \eqref{eq:kappa11} and inserting the result in \eqref{eq:kappa7}, the shell bending rigidity is obtained as a function of the Morse potential depth $\egg$, obtaining

\begin{align}
\kcap\approx 25.66 \egg.
\label{eq:kappa12}
\end{align}

 \begin{figure}[hbt]
  \begin{center}
  \includegraphics[width=\columnwidth]{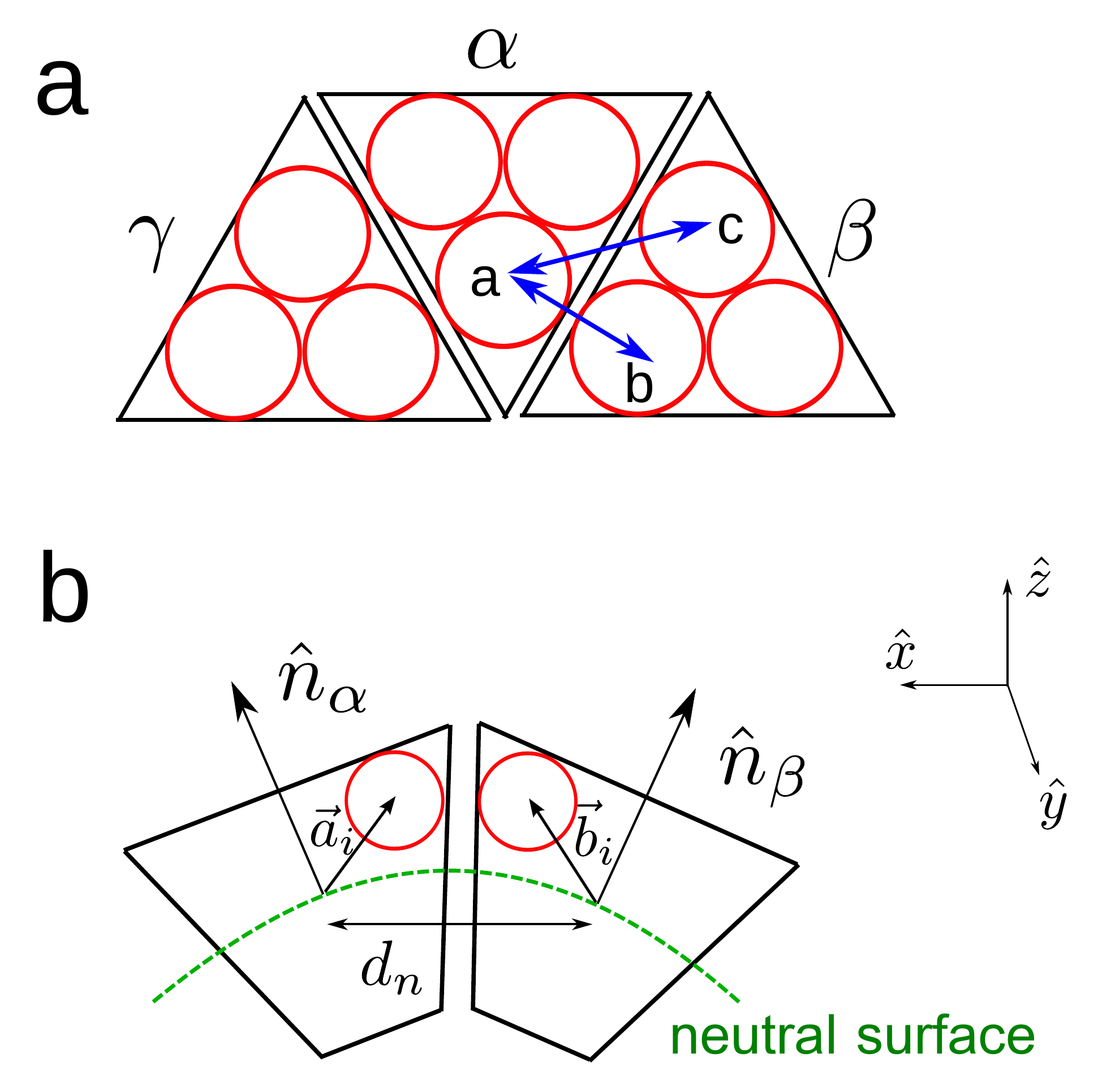}
    \caption{ Subnit organization in the shell.  \textbf{a)} Top view showing the outermost attractive beads in each subunit. For the purpose of estimating the elastic properties of the shell, we consider that all the subunits are organized in such a way that each pseudoatom interacts with two of the pseudoatoms of the neighboring subunits as indicated. \textbf{b)} Lateral view of the subunits in the shell, with the vector positions $\vec{a}_{i}$ and $\vec{b}_{i}$ indicated. The location of the neutral surface is shown by the green dashed line, and $\hat{n}_{\alpha}$ and $\hat{n}_{\beta}$ represent the subunit normal vectors.    }
  \label{fig:sketchelasticity}
  \end{center}
\end{figure}

\subsection{Membrane bending modulus}
The membrane bending modulus is estimated from the height-height fluctuation spectrum. We analyze the fluctuations of a free membrane (\ie without embedded subunits) of size 170x170$\sigma^{2}$. After equilibrating the membrane during $1,500\tau_{0}$, we measure the membrane position for 200 configurations separated by 75$\tau_{0}$. The membrane height $h(\mathbf{x})$ is evaluated from the tail bead positions, and mapped onto a 57x57 grid. Following a standard procedure \cite{Cooke2005,Brandt2011} the undulation modes in real space can be decomposed in modes in Fourier space,

\begin{align}
h(\mathbf{x})= \sum\limits_{\mathbf{q}} h(\mathbf{q}) e^{i \mathbf{q \cdot x}}.
\label{eq:hq}
\end{align}

\noindent where $\mathbf{q}=(q_{x},q_{y})=(n,m) 2\pi / L$. From the equipartition theorem, the fluctuation spectrum reads

\begin{align}
\langle |h(q)|^{2} \rangle= \frac{\kt}{L^{2} \left[ \kappa q^4 + \gamma q^2 \right]},
\label{eq:spectrum}
\end{align}

\noindent where $\gamma$ is the remnant surface tension of the membrane. Even for very small surface tension, in the smaller wave modes the fluctuation spectrum is dominated by tension. For the the analysis we only consider the modes $2\pi/q>5d $, where $d$ is the membrane thickness. Fig. \ref{fig:fluctuations}a shows an example of the fluctuation spectrum for a membrane at $\omega_\text{c}=1.6$. The results are fit to a $q^{-4}$ curve, obtaining the bending rigidity. Fig. \ref{fig:fluctuations}b shows our estimation of the membrane bending modulus as a function of the control parameter $\omega_\text{c}$. Note that at high $\omega_\text{c}>1.65$ our results suggest a slightly lower bending modulus than that obtained by \citet{Cooke2005}.

\section{Subunit and lipid diffusion in the membrane}
\label{sec:diffusion}
We estimate the subunit diffusion constant in the membrane from the subunit mean square displacement  versus time  \cite{Goose2013},

\begin{align}
\langle |\mathbf{r}_{\parallel}(t)-\mathbf{r}_{\parallel}(0)|^{2} \rangle = 4Dt.
\label{eq:msd}
\end{align}

We equilibrate a membrane of size 50x50$\sigma^{2}$ during 1,500$\tau_{0}$, and then sample during 75$\tau_{0}$ with period 0.15$\tau_{0}$. Averaging over 20 independent simulations of a single subunit diffusing on membrane, we obtain a subunit diffusion $D_{\text{sub}}=1.0\sigma^{2}/\tau_{0}$. A similar value is obtained at the subunit concentration used in our simulations (membrane fraction covered by subunits $\approx0.15$), meaning that we have not reached the limit of protein crowding in which diffusion decreases \cite{Goose2013}. Using the same method for the lipids, and averaging over 500 molecules, we obtain $D_{\text{lip}}=0.12\sigma^{2}/\tau_{0}$. The fact that subunits diffuse faster than lipids might be expected, since the subunit pseudoatoms which overlap with the membrane do not interact with the lipids, whereas lipids are subject to much higher friction as they interact with all the neighboring lipids. In biological membranes, however, transmembrane proteins usually diffuse around 3-4 times slower than lipids. This unrealistic fast diffusion was intentionally introduced in the model to speed up the simulation, allowing assembly completion within a tractable simulation time. The characteristic timescale of our simulation is then given by the subunit diffusion, as explained in the main text.

 \begin{figure}[hbt]
  \begin{center}
  \includegraphics[width=\columnwidth]{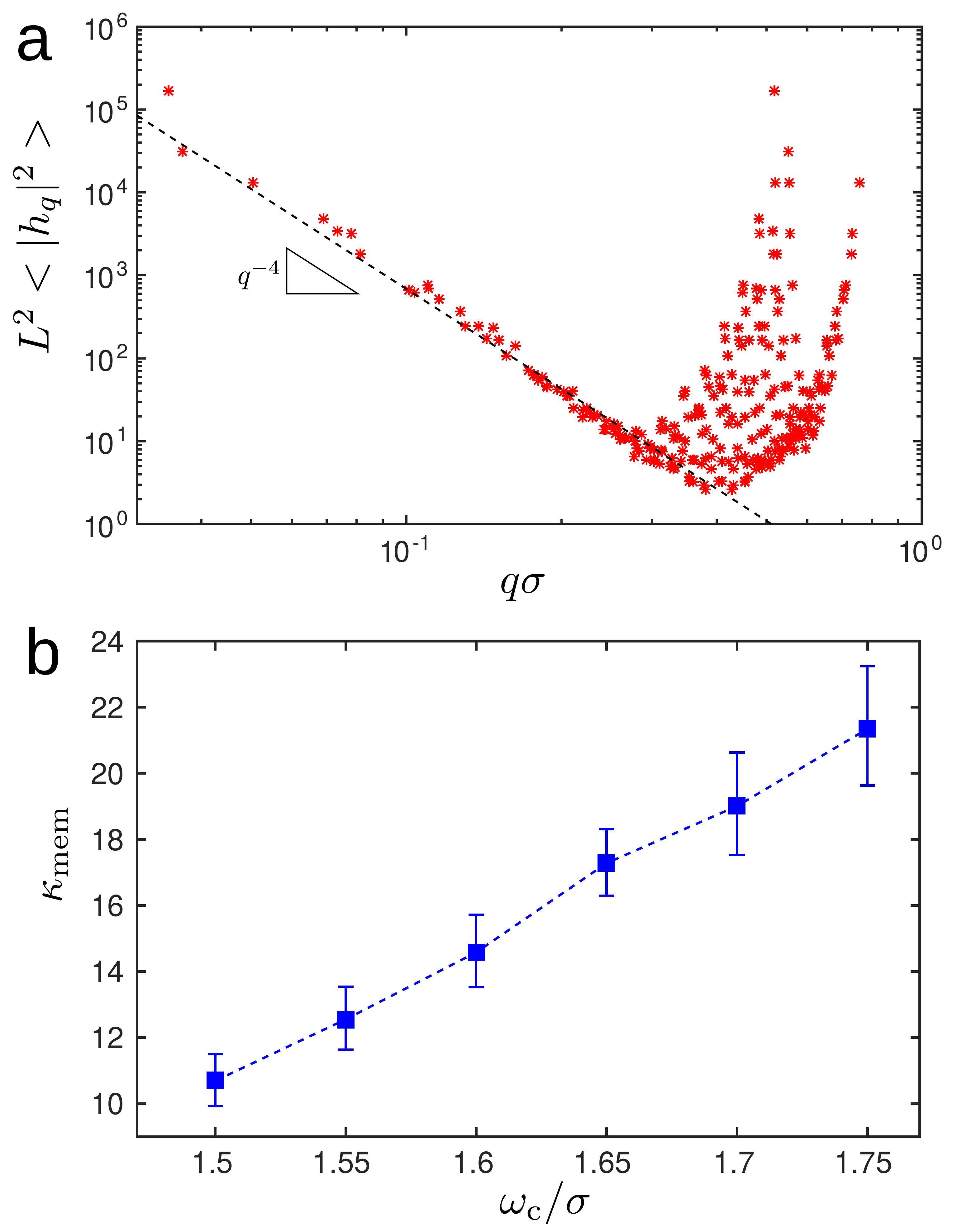}
    \caption{ \textbf{a)} Fluctuation spectrum $<|h(\mathbf{q})|^{2}>$ as a function of the wave mode $q$, for $\omega_\text{c}=1.6$. The dashed line represents the fit curve $q^{-4}$. \textbf{b)} Membrane bending modulus $\kmem$ measured from the fluctuation spectrum as a function of the parameter $\omega_\text{c}$.}
  \label{fig:fluctuations}
  \end{center}
\end{figure}


\end{document}